\documentclass[traditabstract]{aa}  

\usepackage{graphicx}
\usepackage{txfonts}
\usepackage{comment}
\usepackage{enumitem}
\usepackage{supertabular}
\usepackage{longtable,pdflscape}
\usepackage{afterpage}
\usepackage{tikz}
\usepackage{tabu}
\providecommand{\sorthelp}[1]{}

\def\planck{\textit{Planck}}

\newcommand{\ngal}{{N_{\rm gal}}}

\newcommand{\pl}{\left(}
\newcommand{\pr}{\right)}

\def\kms{{\rm km}\:{\rm s}^{-1}}
\def\lesssim{\mathrel{\hbox{\rlap{\hbox{\lower4pt\hbox{$\sim$}}}\hbox{$<$}}}}
\def\gtrsim{\mathrel{\hbox{\rlap{\hbox{\lower4pt\hbox{$\sim$}}}\hbox{$>$}}}}

\def\degree{\mbox{$^{\circ}$}}

\newcommand{\msun}{\hbox{M$_{\odot}$}}

\usepackage{amstext}

\begin{document} 


\title{Velocity dispersion and dynamical mass for 270 galaxy clusters in the \planck\ PSZ1 catalogue}


\titlerunning{ITP MOS spectroscopy and PSZ1 mass scaling relation}

   \author{A.~Ferragamo\inst{1,2,3}
          \and R.~Barrena\inst{1,2} 
          \and J.~A.~Rubi\~{n}o-Mart\'{\i}n\inst{1,2}
          \and A.~Aguado-Barahona\inst{1,2}
          \and A.~Streblyanska   \inst{1,2,4}
          \and D.~Tramonte \inst{5,1,2}
          \and R.~T.~G\'enova-Santos \inst{1,2}
          \and A.~Hempel \inst{6,7}
          \and H.~Lietzen \inst{8}
          }

\institute{Instituto de Astrof\'{\i}sica de Canarias (IAC), C/ V\'{\i}a L\'actea s/n, E-38205 La Laguna, Tenerife, Spain
  \and Universidad de La Laguna, Departamento de Astrof\'{\i}sica, C/ Astrof\'{\i}sico Francisco S\'anchez s/n, E-38206 La Laguna, Tenerife, Spain
  \and  Dipartimento di Fisica, Sapienza Universit\`a di Roma, Piazzale Aldo
Moro, I-00185 Roma, Italy 
\and Gran Telescopio Canarias, La Palma, Spain
\and Purple Mountain Observatory, No. 8 Yuanhua Road, Qixia District, Nanjing 210034, China
\and Universidad Andr\'es Bello, Departemento de Ciencias F\'{\i}sicas, 7591538 Santiago de Chile, Chile
\and Max-Planck Institute for Astronomy, K\"onigstuhl 17, D-69117 Heidelberg, Germany
\and Tartu Observatory, University of Tartu, 61602 T\~{o}ravere, Tartumaa, Estonia \\}

   \date{}

 
\abstract{

We present the velocity dispersion and dynamical mass estimates for 270 galaxy clusters included in the first Planck Sunyaev-Zeldovich (SZ) source catalogue, the PSZ1. %
Part of the results presented here were achieved during a two-year observational program, the ITP, developed at the Roque de los Muchachos Observatory (La Palma, Spain).
In the ITP we carried out a systematic optical follow-up campaign of all the 212 unidentified PSZ1 sources in the northern sky that have a declination above $-15^\circ$ and are without known counterparts at the time of the publication of the catalogue. 
We present for the first time the velocity dispersion and dynamical mass of 58 of these ITP PSZ1 clusters, plus 35 newly discovered clusters that are not associated with the PSZ1 catalogue. 
Using Sloan Digital Sky Survey (SDSS) archival data, we extend this sample, including 212 already confirmed PSZ1 clusters in the northern sky. 

Using a subset of 207 of these galaxy clusters, we constrained the $M_{\rm SZ}$--$M_{\rm dyn}$ scaling relation, finding a mass bias of $(1-B) = 0.83\pm0.07$(stat)$\pm0.02$(sys). We show that this value is consistent with other results in the literature that were obtained with different methods (X-ray, dynamical masses, or weak-lensing mass proxies). 
This result cannot dissolve the tension between primordial cosmic microwave background anisotropies and cluster number counts in the $\Omega_{\rm M}$--$\sigma_8$ plane. 

}


\maketitle
%

%
\section{Introduction}
\label{sec:intro}
According to the bottom-up hierarchical scenario, Galaxy clusters (GCs) are the last bounded structures to form in our Universe (starting from about $z \sim 2$). They reside in the deepest potential wells, which are generated by dark matter (DM) overdensities \citep[e.g.][]{springel05}. 
Galaxy clusters are multi-component structures. In addition to DM, haloes include baryonic matter in different forms and phases \citep[see e.g.][and references therein]{allen11}. In addition to galaxies, cold and hot gas and non-thermal plasma constitute the intra-cluster medium (ICM). This multi-component nature allows us to observe GCs at various wavelengths, taking advantage of the different physical processes involved in each case. For instance, we use X-rays and radio observations to probe the ICM or the visible/IR wavelengths to study the galactic component. Dark matter is usually studied through the deformation of images from the background galaxies. 
These multi-wavelength observations provide complementary information about the cluster physics. 

Galaxy clusters are excellent tracers of the evolution of structures throughout the history of the Universe. Their abundance as a function of redshift and mass, the so-called cluster number counts, is very sensitive to cosmological parameters, and in particular, to the matter density parameter $\Omega_m$ and to the amplitude of the density fluctuations $\sigma_8$ \citep{vikhlinin09, mantz10, planck13_count, planck15_count}.
The accuracy in the cluster mass determination is crucial for the determination of cosmological parameters. To this end, several mass proxies can be defined through scaling relations \citep[see e.g.][and references therein]{Pratt19}. However, each mass proxy has its own biases that are linked to the methods of observation or the assumptions made to calculate it. To understand these biases, the analysis of cosmological simulations is a valuable tool, and the comparison of the masses obtained from different proxies becomes essential.

The Sunyaev-Zeldovich (SZ) effect \citep{SZ72} is a spectral distortion of the cosmic microwave background (CMB), generated by the inverse Compton scattering of the CMB photons off the hot electrons in the ICM. As the brightness temperature of the resulting distortion is redshift independent, the SZ effect has recently become a powerful tool for detecting GCs. 

The ESA\planck{} Planck mission scanned the entire sky in microwaves with the aim of studying primary and secondary anisotropies of the CMB. The observations, which lasted four years, spanned nine bands between 30\,GHz and 857\,GHz. The results of these observations include two catalogues of objects detected by means of their SZ signature: PSZ1 \citep{planck13_cat, planck13_cat_2} and PSZ2 \citep{planck15_cat}. 
These products were the first to provide the possibility of detecting GCs through the SZ effect in a full-sky survey. The catalogues contain 1227 and 1653 objects, respectively, with 937 objects in common. %
The total integrated SZ signal within a circle of radius (the so-called integrated Compton
parameter, $Y_{\rm SZ}$) is closely related to the cluster mass \citep[e.g.][]{daSilva2004}. This $Y_{500}$ observable has been used as the mass proxy by the Planck Collaboration, after calibrating it to X-ray measurements \citep{planck13_count, planck15_count}. 

This paper is the third (and last) in the series of publications associated with the International Time Programme ITP13B-15A, a two-year (August 2013 to July 2014) long-term program in the Canary Islands Observatories dedicated to characterising PSZ1 sources in the northern sky without known optical counterparts at the time of publication of the PSZ1 catalogue. 
The publications were a continuation of the validation efforts carried out in \cite{nostro16}, within the context of the ITP 12-2 program.
Paper I \citep{rafa18} and paper II \citep{rafa20} described the ITP13B-15A program in detail, which included observations of 256 SZ sources with a declination above $ -15^\circ$ (212 of them were previously unknown), finding optical counterparts for 152 SZ sources. This paper (number III in the series) presents all the spectroscopic observations of the program, including velocity dispersion and dynamical mass estimates. 

In section \ref{sec:ITP_SDSS_sample}, we describe the PSZ1-North reference sample summarising the results of the ITP program.
Sections \ref{subsec:POFU_spec} and \ref{sec:dyn_mass} depict how we estimate GC velocity dispersion and dynamical masses, respectively.
In section \ref{sec:ITP_SDSS_sample} we describe the ITP and the SDSS GC samples we used for the cosmological analyses of this work. In sections \ref{subsec:relation2}, we constrain the mass bias parameter, $\left (1- B\right)$, explaining how much the Eddington bias affects our sample and detailing the importance of correcting for it. Finally, in section \ref{sec:comparison}, we compare our results with those from \planck{} Collaboration and those from the literature, using the velocity dispersion and the weak-lensing analysis as mass proxies. We also discuss the implications on the $\sigma_8$ tension with CMB measurements.

Throughout this paper, we define $R_{200}$ and $R_{500}$ as the radius within which the mean cluster density is 200 and 500 times the critical density of the Universe at redshift $z$, respectively. The quantities with the subscripts 200 or 500 have to be considered as evaluated at or within $R_{200}$ and $R_{500}$, respectively.
We assume a flat $\Lambda$CDM cosmology, with $H_0 =100 h$\,km\,s$^{-1}$\,Mpc$^{-1}$, $h = 0.678$,  and $\Omega_{\rm M}= 0.307$ \citep{planck13_par}.

\section{PSZ1-North sample}
\label{sec:ITP_SDSS_sample}

The PSZ1 catalogue \citep{planck13_cat,planck13_cat_2} contains 1227 sources detected by means of their SZ signature in the all-sky maps obtained during the first 15.5 months of \planck\ observations. 
These sources are detected by at least one of the three
\planck\ cluster detection algorithms (MMF1, MMF3, and PwS) with a signal-to-noise ratio (S/N) of $4.5$ or higher. \cite{planck13_cat} describes these three algorithms, the selection, and the validation methods adopted in the construction of the catalogue in detail. 

Following papers I and II, we define the PSZ1-North sample as the 753 PSZ1 sources in the northern hemisphere with a declination $Dec \geq -15\degree$. Of these, 541 were validated by the \planck\ collaboration at the time of publication of the PSZ1 catalogue. The remaining 212 sources were studied within the ITP program. 

\begin{figure*}
\centering
\includegraphics[trim=0.4cm 1.5cm 0.4cm 1.5cm, clip, width=\textwidth]{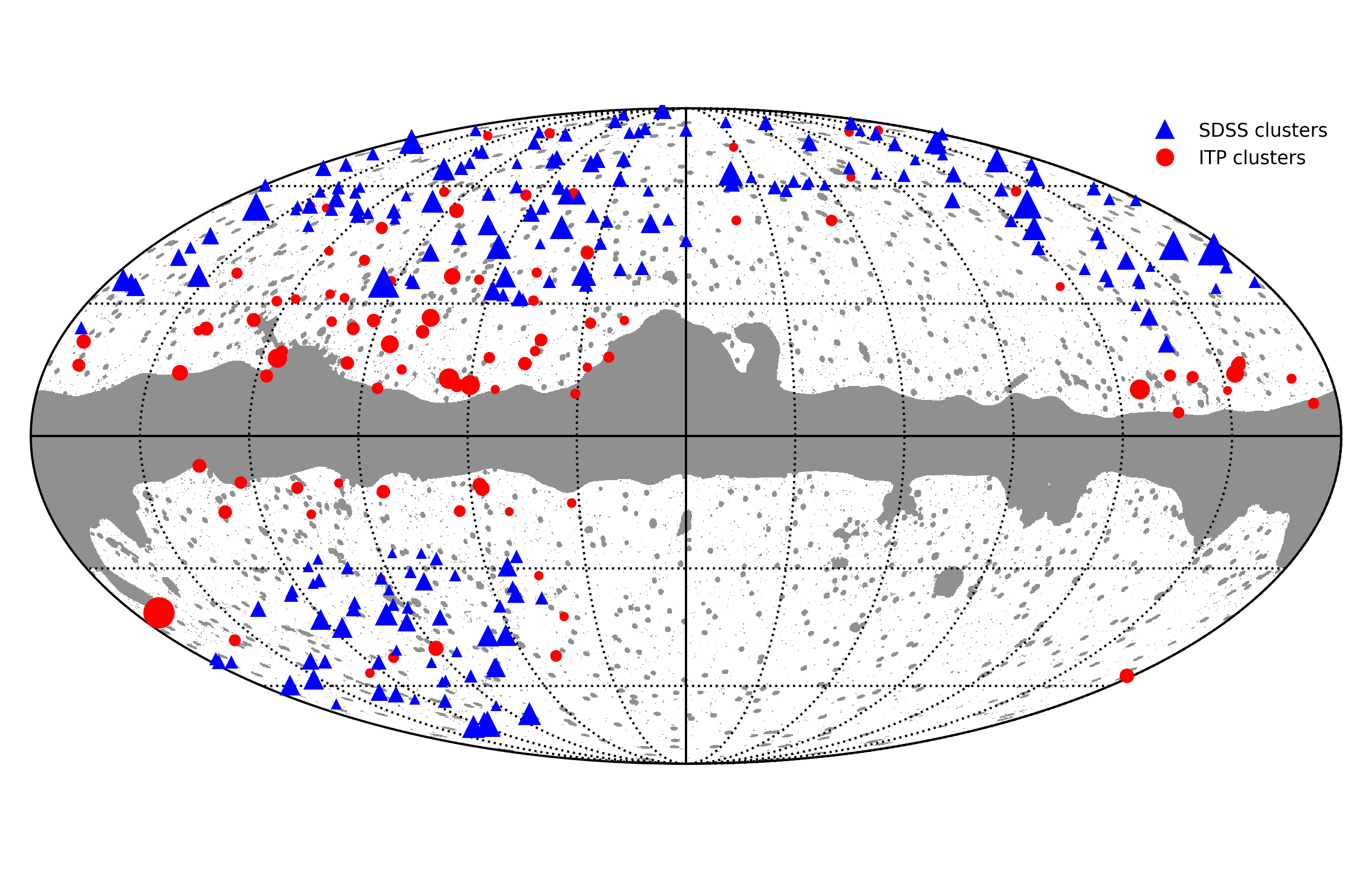}
\caption{Sky distribution of all the PSZ1 clusters studied in this paper. The figure uses a Mollweide projection in Galactic coordinates. The Galactic plane is horizontal and centred at longitude zero. Red circles and blue triangles represent ITP (Sec.~\ref{sub:itp}) and SDSS  (Sec.~\ref{sub:sdss}) clusters, respectively, and the symbol size is proportional to the cluster mass. The shaded grey area represents the union mask used to produce the CMB map in the first \planck\ data release (downloaded from the Planck Legacy Archive). This mask excludes $27\,\%$ of the sky, mostly in the region of the Galactic plane, the Magellanic Clouds, and point sources. }
\label{fig:fullsample}
\end{figure*}

\begin{table}
\centering
\caption{Summary of the data sets.}
\begin{tabular}{c c c|c}
\hline \hline
\noalign{\smallskip}
Data set    & PSZ1-North  & Beyond PSZ1 & Scaling relation\\
 & & (see Sect.~\ref{sub:itp}) &  (see Sect.~\ref{sec:subsample}) \\
\hline
\noalign{\smallskip}
ITP   &  58 & 35 &  58 \\
SDSS  & 212 &  0 & 149 \\
\hline
Total & 270 & 35 & 207 \\
\hline \hline
\end{tabular}
\label{tab:sample}
\end{table}

\subsection{ITP program}
\label{sub:itp}
\cite{nostro16}, \cite{rafa18}, and \cite{rafa20} have described our optical follow-up of the unknown PSZ1 sources in the northern sky that we carried out between the second semester of 2012 and first semester of 2015 by means of two International Time Projects (ITP 12-2 and ITP13-08, hereafter ITP).
Table~3 in \cite{rafa20} presented the summary information of the full program, where we observed all the 212 PSZ1-North sources for which no counterparts were known, plus 44 of the already validated sources. 

Papers I and II have presented the complete imaging results of the program and partial spectroscopic results, including the mean spectroscopic redshifts to the clusters, as well as the redshift of the BCG, when available. This paper III presents for the first time the velocity dispersion and dynamical mass estimates for s with at least $\ngal \text{seven}$ spectroscopic members. In total, 58 PSZ1-North clusters were characterised (see Table~\ref{tab:sample})

The majority of these results were obtained using multi-object spectroscopy (MOS), although for a few cases and when the MOS mode was not available, we used a combination of long-slit measurements. 
As described in previous papers, all the spectroscopic data for the ITP program have been acquired using three instruments at the Observatorio del Roque de Los Muchachos (ORM), located on the island of La Palma (Spain). 
OSIRIS at the 10.4 m Gran Telescopio Canarias (GTC) and DOLORES at the 3.5 m Telescopio Nazionale Galileo (TNG) were mainly used to obtain MOS, while ACAM at the 4.2 m William Herschel Telescope (WHT) provided some additional spectra that were retrieved through a long-slit setup.
A detailed description of the instrument characteristics and the configuration adopted for these measurements has been presented in papers I and II.

When the MOS technique was used with OSIRIS and DOLORES, we observed each field with a single mask containing between 45 and 60 slitlets on average. In this way, we were able to maximize the number of redshifts per observation, obtaining a median number of galaxy members of about 20 (see details in paper II). 

Table~\ref{table:ITP_masses_flag1} in the appendix contains the full list of the 58 ITP objects within PSZ1-North that we studied in this paper. The first, second, and third columns identify the index number, the \planck\ name of each cluster in the PSZ1 catalogue, and the signal-to-noise ratio of the SZ detection. Columns 4 and 5 indicate the equatorial coordinates (J2000) of the cluster optical centre. Column 6 indicates the distance from the optical centre to the nominal \planck\ coordinates. 
Columns 7 and 8 contain our spectroscopic redshift and the number of cluster members with spectroscopic measurements, respectively. Columns 9 to 11 contain the velocity dispersion, the dynamical mass, and the SZ mass estimates. The description of the process with which these numbers were derived is discussed in Sections~\ref{subsec:POFU_spec} and \ref{sec:dyn_mass}. Finally, columns 12 and 13 indicate whether the cluster was used to constrain the scaling relation and if it was part of the \planck\ cosmological sample (hereafter, PlCS), as defined in \cite{planck13_cat}.

During this program, we also characterised 35 bona fide GCs that have not been associated with the SZ signal measured by \planck\, either because they lie at a distance of more than $5\arcmin$ from the nominal \planck\ pointing or because their velocity dispersion is too low. These objects were labelled {\tt Flag}$=3$ in our validation process \citep[see details in ][]{rafa18, rafa20}. These 35 clusters are listed in the column "Beyond PSZ1" in Table~\ref{tab:sample} and are presented in Table~\ref{table:ITP_masses_flag3}, using the same format as in Table~\ref{table:ITP_masses_flag1}.

The location of all these ITP clusters is shown in Figure~\ref{fig:fullsample} in red. We obtained more than $2000$ individual redshift measurements in this work. The mean number of galaxy members for these ITP clusters is $\ngal = 15$. 
All the spectra associated with these observations that we present here for the first time will be published online and are included in the VO database. 

\subsection{SDSS archival data}
\label{sub:sdss}
With the aim of characterising the full PSZ1-North sample, we decided to enlarge our ITP sample by searching in the SDSS Data Release 14 \citep{aihara11_sdss} for additional spectroscopic information. 
For each of the $401$ PSZ1 sources within the SDSS footprint, we retrieved all spectroscopic redshifts within $15\arcmin$ radius around the nominal \planck\ coordinates. We used the position of the brightest galaxy in the radial velocity catalogue as the cluster centre, within a range of $\pm 2500\,\kms$ around the \planck\ validation redshift. In most cases, this position corresponded to the BCG. If not, after a visual inspection of SDSS RGB images, we selected the centre as the position of a clear BCG (not observed spectroscopically). Alternatively, we used the cluster members mean coordinates as the optical centre.

For each individual PSZ1 field, we analysed the retrieved set of spectroscopic redshifts in the same way as we treated the clusters that were observed during the ITP program. However, due to the lack of (enough number of) spectra in some fields, it was only possible to characterise the spectroscopic properties for 212 regions. As in the previous case, we only considered the clusters for which we had  at least $\ngal \text{seven}$ spectroscopic members.

Table~\ref{table:SDSS_masses} in the appendix contains the final list of PSZ1-North clusters with SDSS information that is discussed in this paper. The table format is identical to that of Table~\ref{table:ITP_masses_flag1}. 
The location of these 212 clusters is also shown in Figure~\ref{fig:fullsample} in blue. We used $\sim 10000$ spectra in this analysis. The mean number of galaxy members for these clusters is $\ngal = 35$. 

Finally, the joint ITP and SDSS sample consists of 270 GCs, 226 of which are also included in the PSZ2 catalogue. In our sample, the percentage of PSZ1 clusters that are also detected in the PSZ2 is therefore $85\%$. This number is slightly higher than the same fraction computed for the whole PSZ1 catalogue, $76\%$ \citep[see table 10 in][]{planck15_cat}.


\section{Velocity dispersion estimates}
\label{subsec:POFU_spec}
 
For each , the final product of either our ITP observations or the SDSS archival data search is a catalogue of radial velocities of galaxies within a certain field. This means that the cluster is detected as an over-density in the radial velocity distribution of at least four galaxies in a range of $\pm 5000\,\kms$ around the velocity of the BCG. 
The membership selection is based on the galaxy position in the 2D projected phase space $(r, cz)$, where $r$ is the projected distance from the cluster centre, and $cz$ is the galaxy line-of-sight (LoS) velocity. To minimize the presence of interlopers, we defined a region of membership by performing a cut at $r=2.5$\,Mpc (at the redshift of the cluster) and an iterative $2.5-\sigma$ clipping in the $cz$ coordinate, taking into account the radial profile of the expected velocity dispersion \citep{Mamon10}. The first criterion restricts our galaxy sample to a distance between 1.5 and 2 $R_{200}$ (depending on the cluster mass and redshift) and in this way limits the contamination of interlopers at large distances from the centre of the clusters. 

\begin{figure}
\centering
\includegraphics[trim=0.5cm 0.3cm 0.5cm 1.3cm, clip,width=0.49\textwidth]{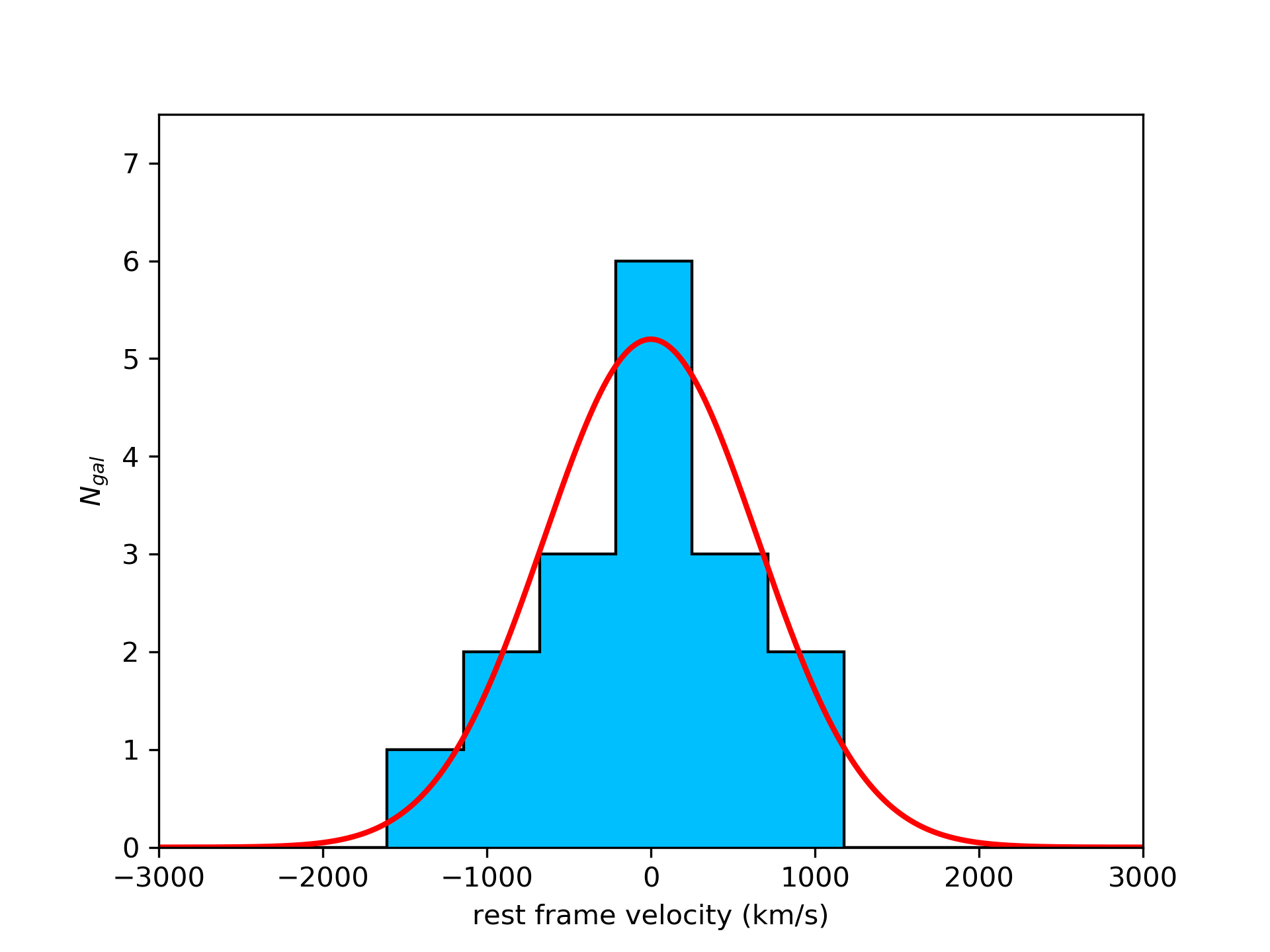}
\includegraphics[trim=0.5cm 0.3cm 0.5cm 1cm, clip,width=0.49\textwidth]{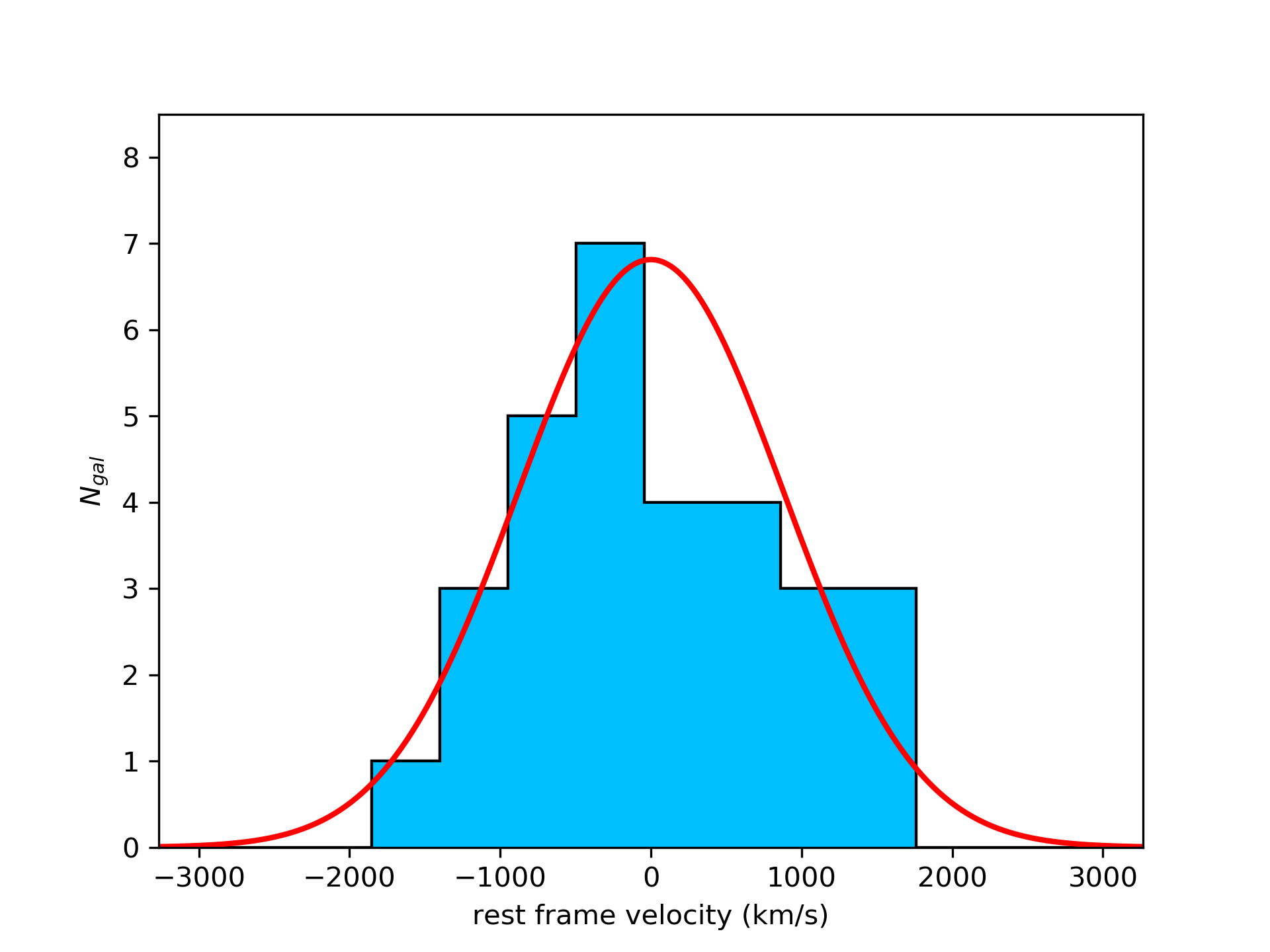}
\caption{Two examples of the velocity distribution of cluster members: PSZ1 G103.94+25.81 (top) and PSZ1 G123.55?10.34 (bottom). Both clusters have been observed within the ITP program. The histograms contain 17 and 30 members, respectively. In both cases, the red line corresponds to a Gaussian distribution centred in the mean cluster velocity, and with $\sigma$ equal to the estimated velocity dispersion $\sigma_v$. }
\label{fig:example_vel_histo}
\end{figure}

\begin{figure}
\centering
\includegraphics[trim=0.5cm 0.4cm 0.5cm 0.5cm, clip, width=0.49\textwidth]{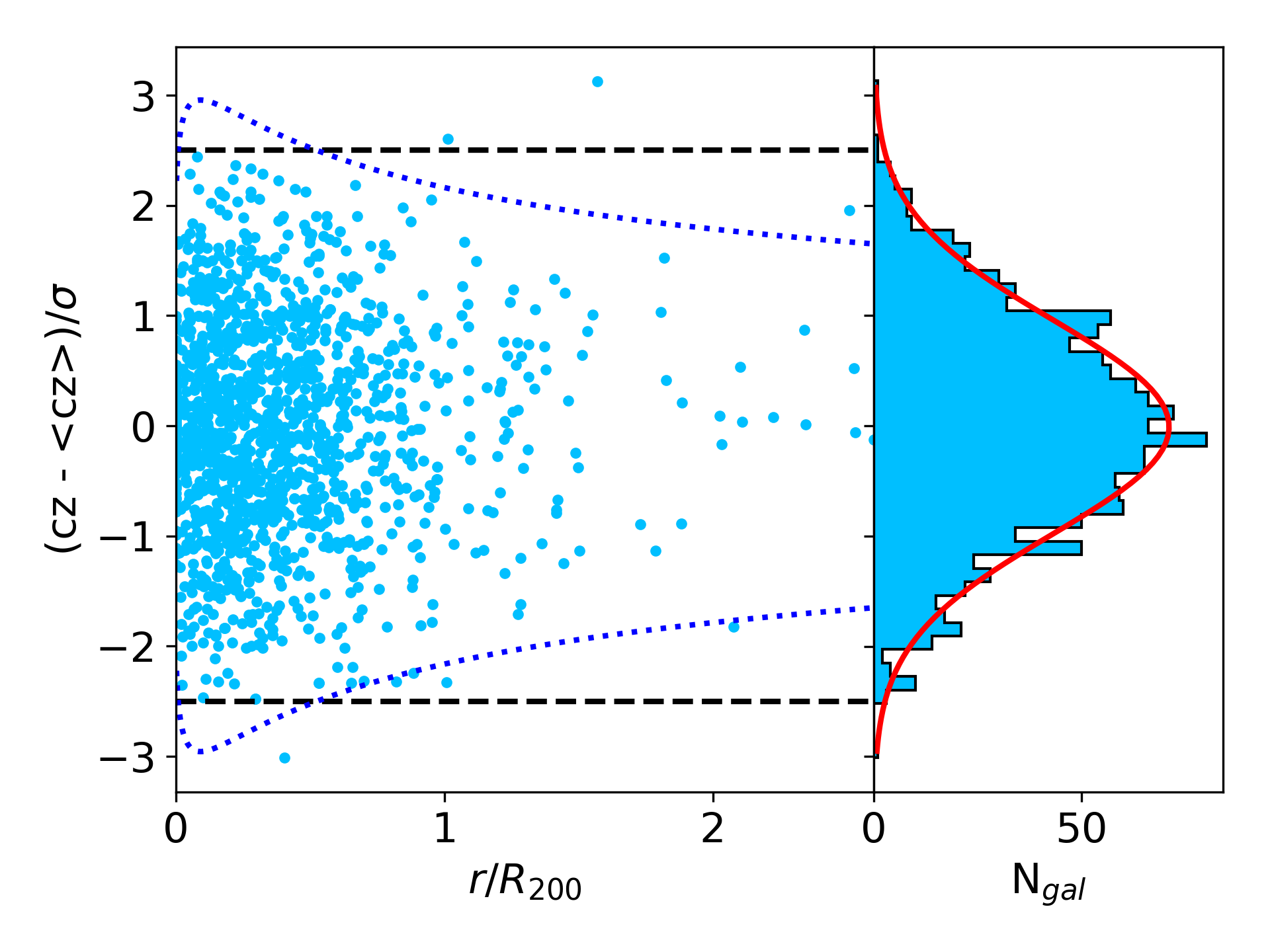}
\includegraphics[trim=0.5cm 0.4cm 0.5cm 0.5cm, clip, width=0.49\textwidth]{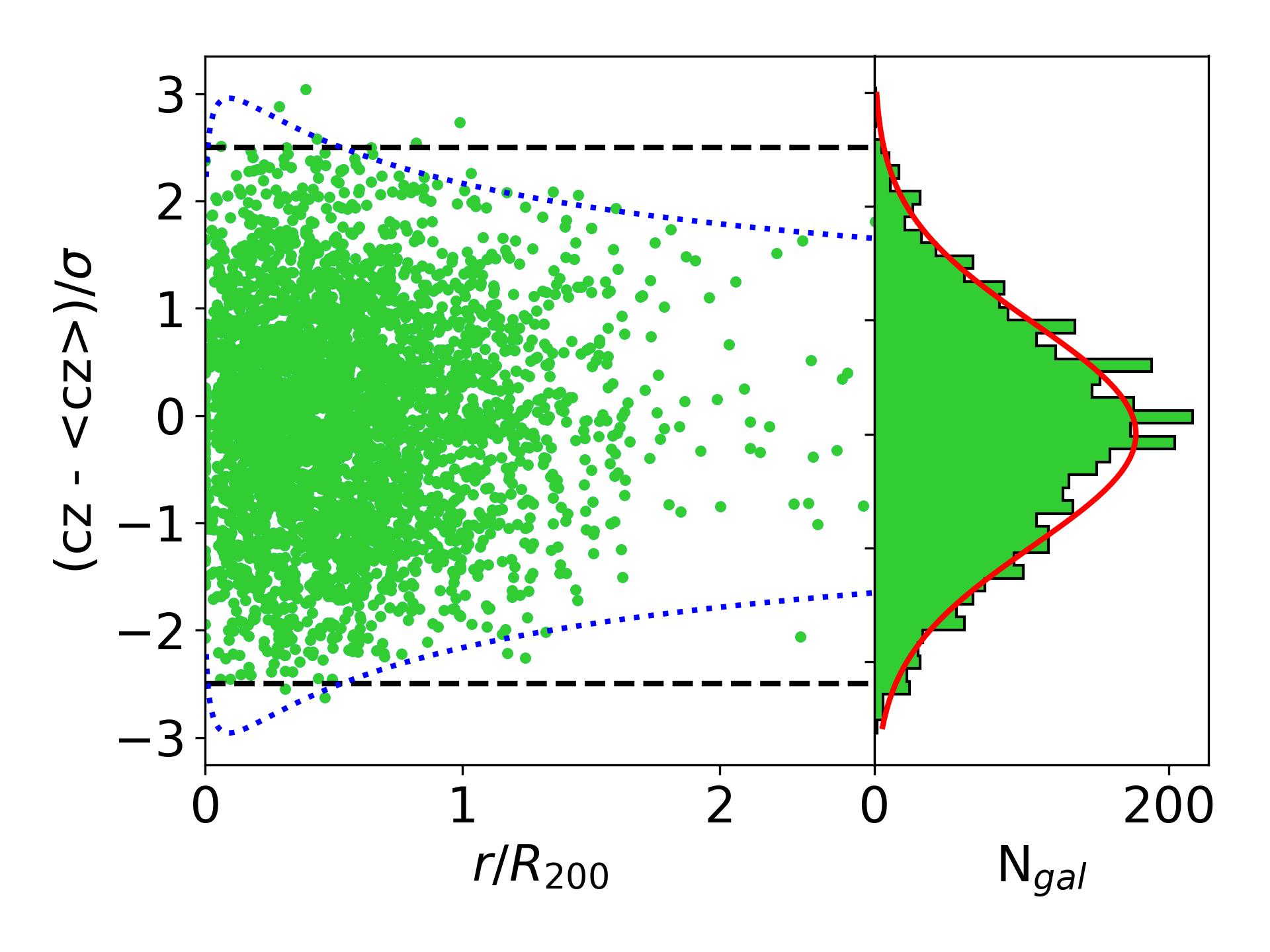}
\caption{Projected phase-space and velocity histogram of the stack of all ITP clusters members (top) and all SDSS cluster members (bottom). In both cases, velocities are normalised to the cluster velocity dispersion. The dashed black lines indicate the $2.5-\sigma$ clipping, and the dotted blue lines show the same clipping, but take the velocity dispersion radial profile as shown in \cite{Mamon10} into account. The red line in the right panel represents a Gaussian fit of the stacked velocity histogram normalised to the total number of members.}
\label{fig:PSZ1_hist}
\end{figure}

Figure~\ref{fig:example_vel_histo} shows two examples of the final velocity histogram of cluster members for two ITP clusters. 
Figure~\ref{fig:PSZ1_hist} shows the 2D projected phase-space and the velocity distribution of the stack of all cluster members selected using the procedure explained above for the ITP sub-sample (1432 clusters) and for the SDSS sub-sample (3879 clusters). 

After this member selection was concluded, we estimated the velocity dispersion using the gapper estimator \citep{gapper}. We followed the procedure outlined in \cite{Ferragamo20}, and we corrected this velocity dispersion estimate by taking the statistical bias introduced by the under-sampling and sigma clipping into account, as well as the physical bias due to the aperture radius. 
Our velocity dispersion results are listed in column 9 of Tables~\ref{table:ITP_masses_flag1}, \ref{table:SDSS_masses}, and \ref{table:ITP_masses_flag3}.

\section{Dynamical mass estimates}
\label{sec:dyn_mass}

We computed the dynamical mass $M_{200}^{\rm dyn}$ by using the bias-corrected mass estimator described in \cite{Ferragamo20}. This estimator is based on the AGN fit of the \cite{munari13} scaling relation, given by
\begin{equation}
\frac{\sigma}{\mathrm{km\ s^{-1}}} =
A\left[\frac{h(z)\ M_{200}}{10^{15}\ \mathrm{M}_{\odot}}\right]^{\alpha},
\label{eq:smg}
\end{equation}
where $A=1177.0$ and $\alpha = 0.364$.

Following prescriptions described in \cite{Ferragamo20}, we corrected
for the aperture radius bias. The physical bias correction due to the fraction of massive galaxies (which tends to reduce the mass of the clusters by $5\,\%$ at most) cannot be evaluated in this work because it requires a well-constrained luminosity function. This cannot be obtained because the clusters we analysed have only a few spectroscopic members.

It follows that the dynamical masses reported here are to be considered as a lower limit to the true mass. The derived mass bias is therefore to be considered as an upper limit. In order to compare our results with the SZ flux ($Y_{500}^{\rm SZ}$) and SZ mass ($M_{500}^{\rm SZ}$) values derived by the \planck\ Collaboration, we need to rescale our values from $M_{200}$ to $M_{500}$. 

To do this, we assumed an NFW density profile \citep{NFW},
\begin{equation}
\rho(r) = \frac{\rho_s}{\left( r/r_s \right)\left( 1+ r/r_s \right)^2}
\label{eq:NFW_profile}
,\end{equation}
where $\rho_s$ and $r_s$ are the characteristic density and the scale radius, respectively. These parameters are related to $R_{200}$ and $\rho_{200} \equiv 200 \times \rho_c$, where $\rho_c$ is the critical density, through the concentration parameter $c_{200}$, which is defined as
\begin{align} 
\label{eq:c_def}
c_{200} \equiv R_{200}/r_s, \\
\rho_s = \frac{\rho_{200} c_{200}^3 g(c)}{3,}
\end{align}
where $g(x) = 1/ \left[ \ln (1+x)-x/(1+x) \right]$. The mass enclosed within a spherical over-density of radius $r_\Delta$ can be written as
\begin{equation}
M(r_\Delta) = \frac{g(c_{200})}{g(c_{200}\;r_\Delta/R_{200})}M_{200},
\label{eq:M_NFW}
\end{equation}
where $\Delta$ is the new density contrast at which we wish to obtain the mass (in our case, $\Delta = 500$).
Finally, we selected a mass-concentration relation to scale our masses \citep{lokas01}. Following the procedure in \cite{Komatsu11_wmap7}, we chose the \cite{duffy08} relation, which is constrained using N-body simulations and WMAP 5-year cosmological parameters,
\begin{equation}
c_{200} = \frac{5.71}{\left( 1+z \right)^{0.47}} \left( \frac{M_{200}}{2 \times 10^{12} h^{-1} \  \rm{M_\odot}} \right)^{-0.084}.
\label{eq:duffy_MC}
\end{equation}

\section{Selection of a representative sub-sample }
\label{sec:subsample}

Here we describe the sub-sample of selected 207 PSZ1-North clusters that we used to characterise the scaling relation $M_{500}^{\rm SZ}$--$M_{500}^{\rm dyn}$. This sub-sample contains all the 58 ITP objects and 149 SDSS clusters (see last column in Table~\ref{tab:sample}) that were selected as explained below. This sub-sample represents $27.5$\,\% of the full PSZ1-North catalogue. 

\subsection{ ITP sub-sample}
Although our observational program has resulted in more confirmed clusters, we considered only the most reliable ones that had an unambiguous association with the \planck\ SZ signal to characterise the scaling relation. Because the measured SZ signal is the total integrated electron pressure along the line of sight, we cannot separate the fraction of the $Y_{500}$ signal that corresponds to each cluster  in the case of multiple cluster counterparts inside the same
pointing. We therefore discarded these multiple counterparts for this analysis. In practice, this does not affect the ITP sample, and we kept all 58 ITP clusters within PSZ1-North, as listed in Table~\ref{table:ITP_masses_flag1} (see the appendix). The last column in the table indicates the objects that are included in the PlCS. By construction, this number is very small (only 2) because the ITP sample includes PSZ1-North objects with unknown counterparts at the time of publication of the catalogue. 

\begin{figure*}
\begin{center}
\includegraphics[trim=0.4cm 0.4cm 0.4cm 0.4cm, clip, width=0.49\textwidth]{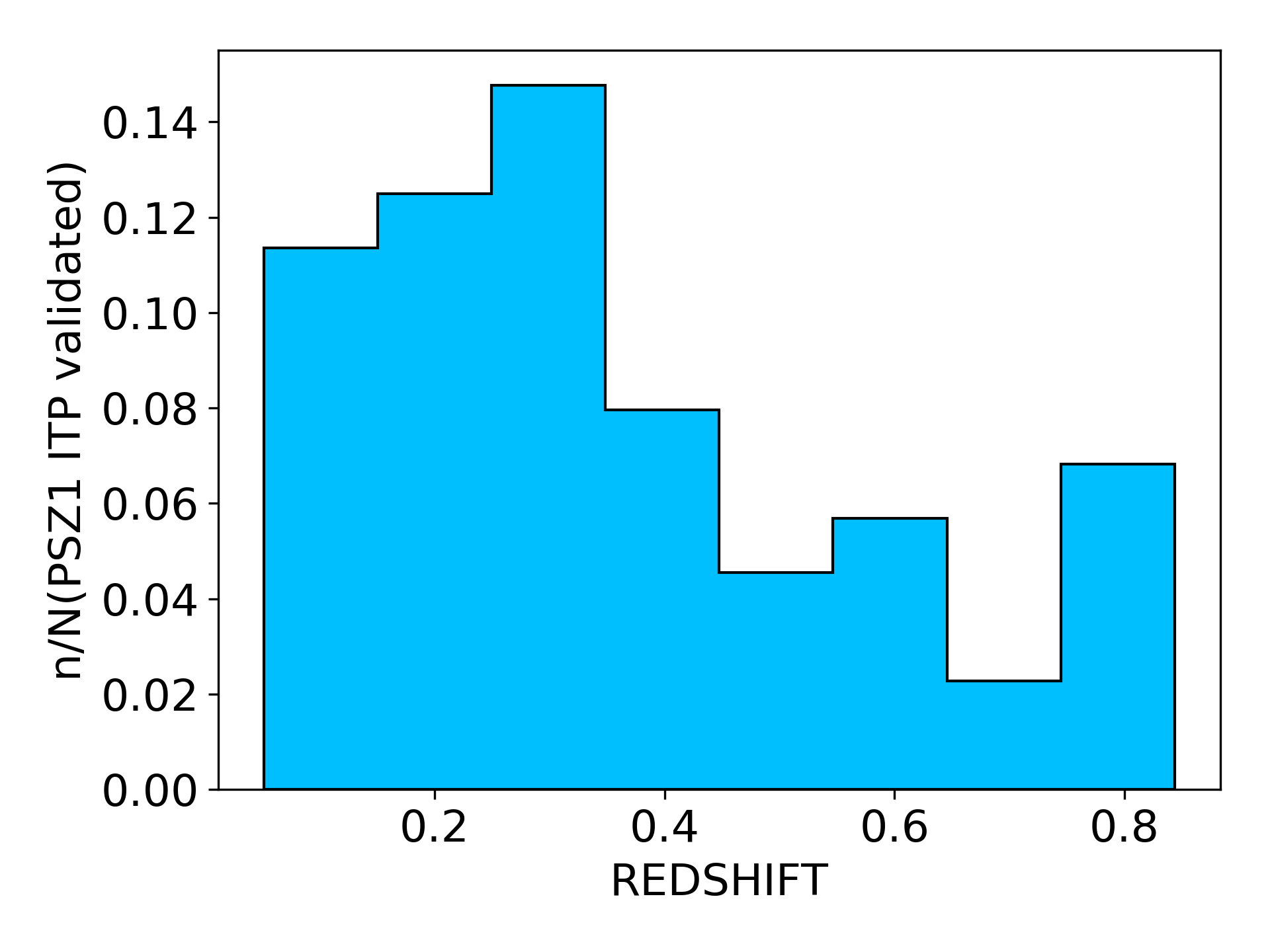}
\includegraphics[trim=0.4cm 0.4cm 0.4cm 0.4cm, clip, width=0.49\textwidth]{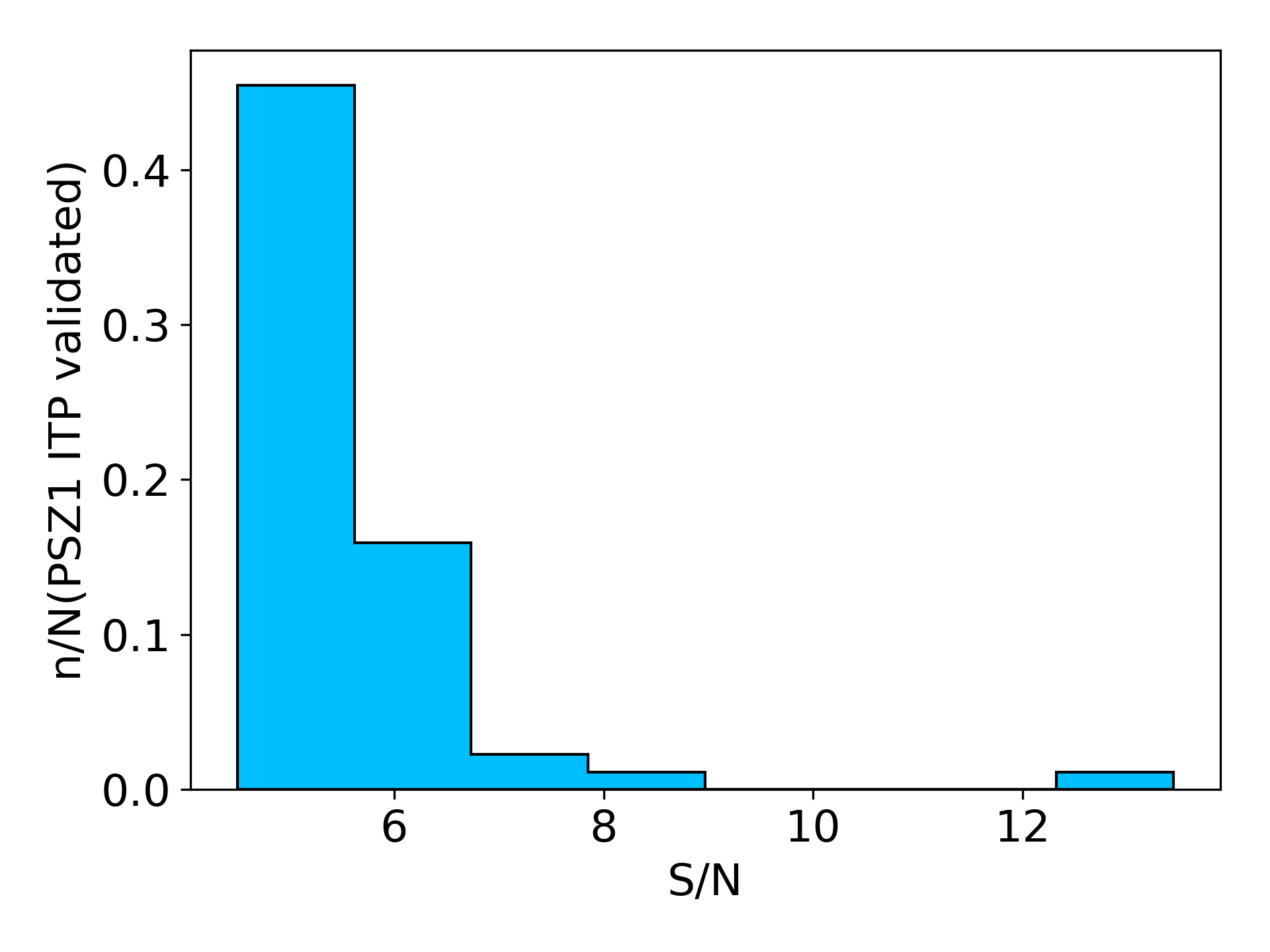}
\end{center}
\caption{Number of ITP clusters as a function of redshift (left panel) and the $S/N$ of the SZ detection (right panel) we used, normalised to 88, which is the total number of ITP-validated clusters \citep[for details, see ][]{rafa20}. }
 \label{fig:z_snr_ITP_hists}
\end{figure*}

Figure~\ref{fig:z_snr_ITP_hists} shows the distribution of these 58 objects as a function of redshift (left panel) and the S/N of the SZ detection (right panel). The median redshift of this sub-sample is $z=0.31$, and the median S/N is $4.9$. 

\subsection{SDSS sub-sample}
From the full list of 212 SDSS objects discussed in Sec.~\ref{sub:sdss}, we only considered the 149 SZ clusters with a single optical counterpart at the same redshift as the validation counterpart reported in the PSZ1 catalogue. These clusters lie within $5\arcmin$ from the nominal \planck\ pointing. 
Although these prescriptions ($\ngal \ge 7$ and a single counterpart) leave us with only $37\%$ of all the 401 PSZ1 clusters within the SDSS footprint, they guarantee a clean  SDSS sample. A detailed and systematic study of the optical counterparts of previously validated clusters of the PSZ1 catalogue is beyond the scope of this work. 

\begin{figure*}
\begin{center}
\includegraphics[trim=0.4cm 0.4cm 0.4cm 0.4cm, clip, width=0.49\textwidth]{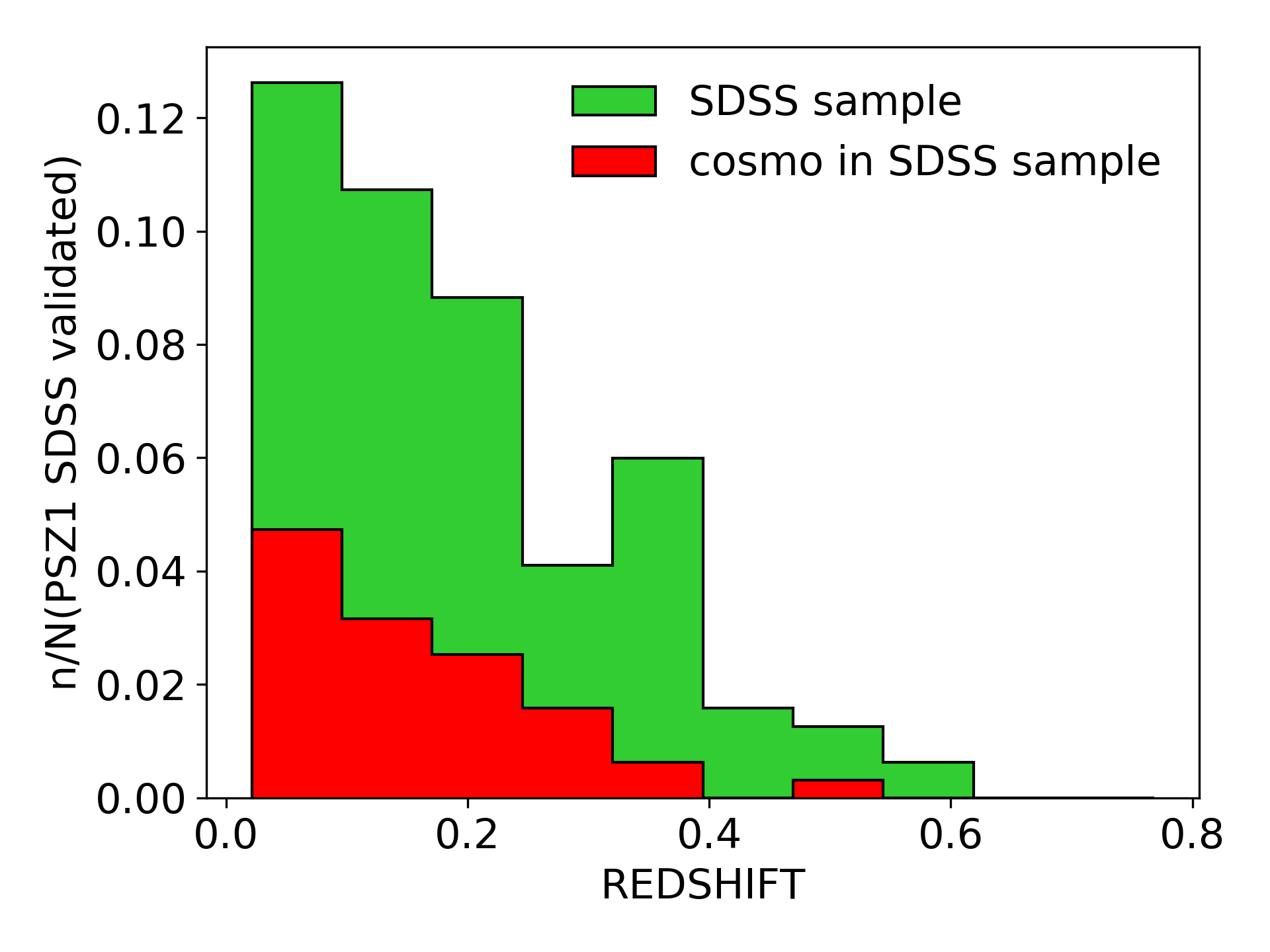}
\includegraphics[trim=0.4cm 0.4cm 0.4cm 0.4cm, clip, width=0.49\textwidth]{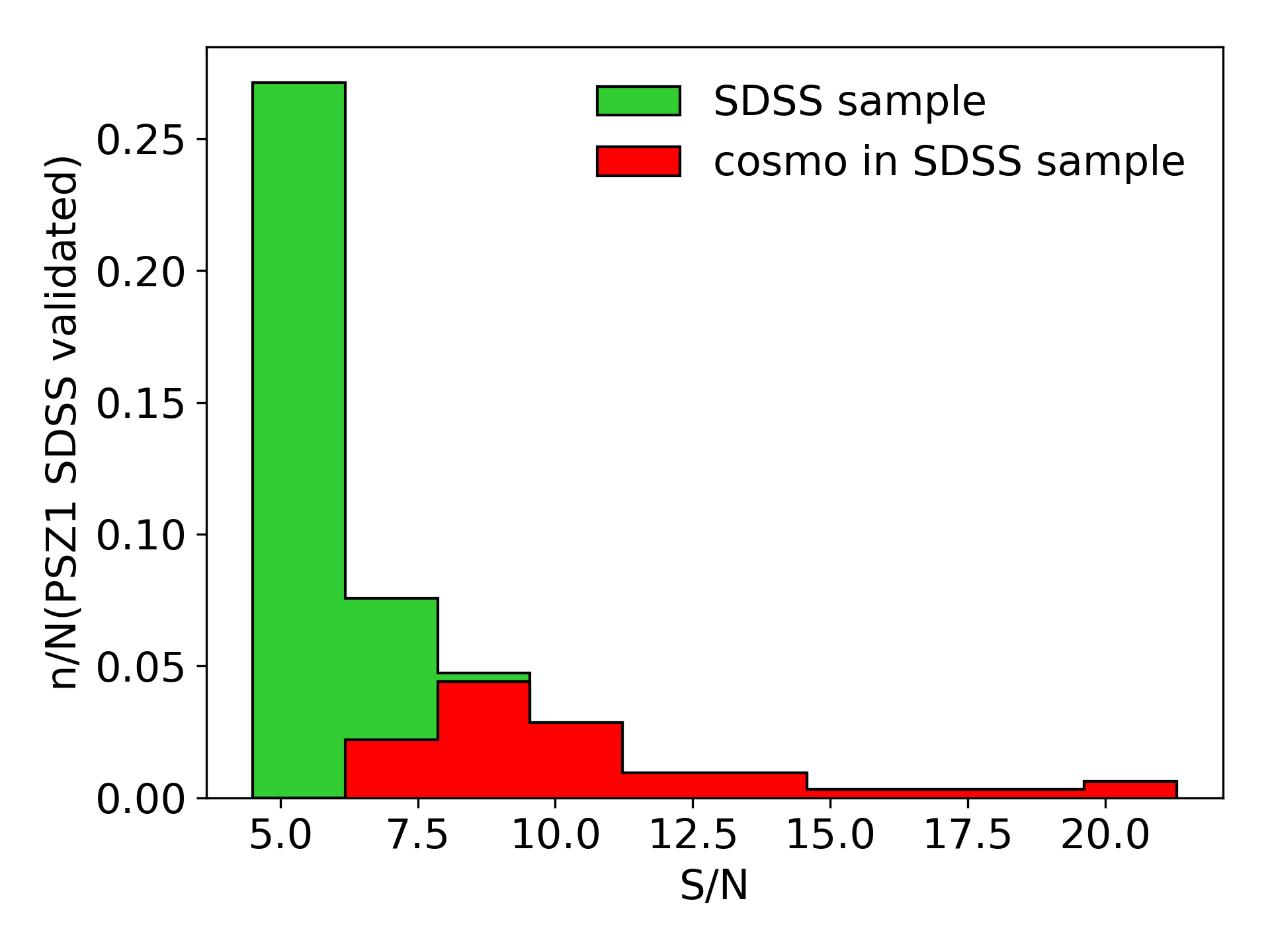}
\end{center}
\caption{Number of SDSS clusters as a function of redshift (left panel) and SZ $S/N$ (right panel) we used, normalised to the total number of PSZ1 sources within the SDSS footprint (green). Red histograms correspond to the subsample of SDSS clusters within the \planck{} cosmological sample.}
 \label{fig:z_snr_SDSS_hists}
\end{figure*}

Figures \ref{fig:z_snr_ITP_hists} and \ref{fig:z_snr_SDSS_hists} show the number of GCs in the ITP and our SDSS sample (green) as a function of the redshift (left panel) and of the $S/N$ (right panel), respectively. The number of PlCS clusters contained in our SDSS sample is plotted in red. Most of the selected clusters are low-redshift systems with a low S/N. The median redshift and $S/N$ are $0.17$ and $5.74$, respectively.
It is also interesting to note that although the percentage of validated PSZ1 clusters comprised in SDSS is $25.8\%$, only $18.6\%$ of the clusters are in the PlCS. However, 42 (28.2\,\%) of the 149 SDSS clusters that we selected are part of the PlCS, with a median redshift equal to 0.11 and a median $S/N$ of 9.39.
In our analysis below, we explore whether the measured mass bias differs for clusters within or outside the cosmological sample. 

In Table \ref{table:SDSS_masses} (see appendix) we list the velocity dispersion and the dynamical and the SZ mass for these $149$ clusters selected from the SDSS sample. The selected objects are marked with a checkmark in the column "scaling". We also indicate the clusters that are included in the cosmological sample. 

\section{$M^{\rm dyn}$--$M^{\rm SZ}$ relation}
\label{subsec:relation2}

Here we describe the results of the $M^{\rm dyn}_{500}$--$M^{\rm SZ}_{500}$ scaling relation using the PSZ1 GC samples selected in the previous section. 
Following the definition given in \cite{planck13_count}, the bias of SZ-derived masses is given by
\begin{equation}
M_{500}^{\rm SZ} \equiv \left( 1-b_{\rm SZ} \right) M_{500}^{\emph{true}},
\label{eq:(1-bsz)}
\end{equation}
where $M_{500}^{\rm SZ}$ is the cluster SZ mass. This is estimated following the prescriptions in \cite{planck13_cat} using the equation 
\begin{equation}
E^{-2/3}(z) \left[\frac{D_{A}^2\ Y_{500}^{\rm SZ}}{10^{-4}\ \rm Mpc^2} \right] = 10^{-0.19 \pm 0.02} \left[ \frac{M_{500}^{\rm SZ}}{6 \times 10^{14}\ \rm M_{\odot}} \right] ^{1.79 \pm 0.08},
\label{eq:Y500_MYx_1}
\end{equation}
which is calibrated using X-ray observations of 71 clusters in the PSZ1 cosmological sample. 

In general, because we cannot account for all the possible biases in the dynamical mass estimation, we defined a global bias of the dynamical mass as
\begin{equation}
M_{500}^{\rm dyn} \equiv \left( 1-b_{\rm dyn} \right) M_{500}^{\emph{true}}.
\label{eq:(1-bdy)}
\end{equation}
Combining equations~\ref{eq:(1-bsz)} and \ref{eq:(1-bdy)}, we obtain the $M_{\rm dyn}-M_{\rm SZ}$ relation as 
\begin{equation}
M_{500}^{\rm SZ} = \left( 1-B \right) M_{500}^{\rm dyn},
\label{eq:(1-B)}
\end{equation}
where the mass bias $(1-B)$ is defined as the ratio of SZ and dynamical bias,
\begin{equation}
    (1-B) \equiv \frac{\left( 1-b_{\rm SZ} \right)}{\left( 1-b_{\rm dyn} \right)}.
\end{equation}

\subsection{Linear regression method}
We fit for the relation in equation~\ref{eq:(1-B)} using a linear regression in logarithmic space because the high noise level of our dynamical mass estimates, with typical 
relative errors of $\sim 30$ \%, could affect the estimation of the fitted parameters. In some cases, we also explored the possibility of a mass dependence by letting the exponent vary freely in the relation, fitting for the slope and intercept in
\begin{equation}
\ln M_{500}^{\rm SZ} = \ln \left( 1-B \right) + \alpha \ln M_{500}^{\rm dyn}.
\label{eq:fit_log}
\end{equation}
This problem requires carrying out a linear regression with errors in both axes, and accounting for intrinsic scatter. In order to select the most appropriate method for this problem, we tested five methods that are commonly used in literature, namely the orthogonal distance regression (ODR, implemented in the python \textit{scipy.odr} package), the Nukers \citep{Tremaine02}, the maximum likelihood estimator with uniform prior (MLEU), the bivariate correlated errors and intrinsic scatter \citep[BCES,][]{akritas96}, and the complete maximum likelihood estimator \citep[CMLE,][]{kelly07}, on realistic mock realisations of PSZ1 sample. In a companion paper associated with the characterisation of the PSZ2 sample \citep{aguado20}, we show that although all these five methods produced biased results in the recovered slope because of the high noise levels in our sample, the Nukers and the ODR present the lowest bias. For this reason, we adopted ODR as the reference regression method, and we tested it using simulations with realistic noise levels mimicking those present in our sample. See Appendix B in \cite{aguado20} for a detailed description of the other estimators.

\begin{figure}
        \centering
        \includegraphics[trim=0.5cm 0.4cm 0.5cm 0.5cm, clip, width=0.49\textwidth]{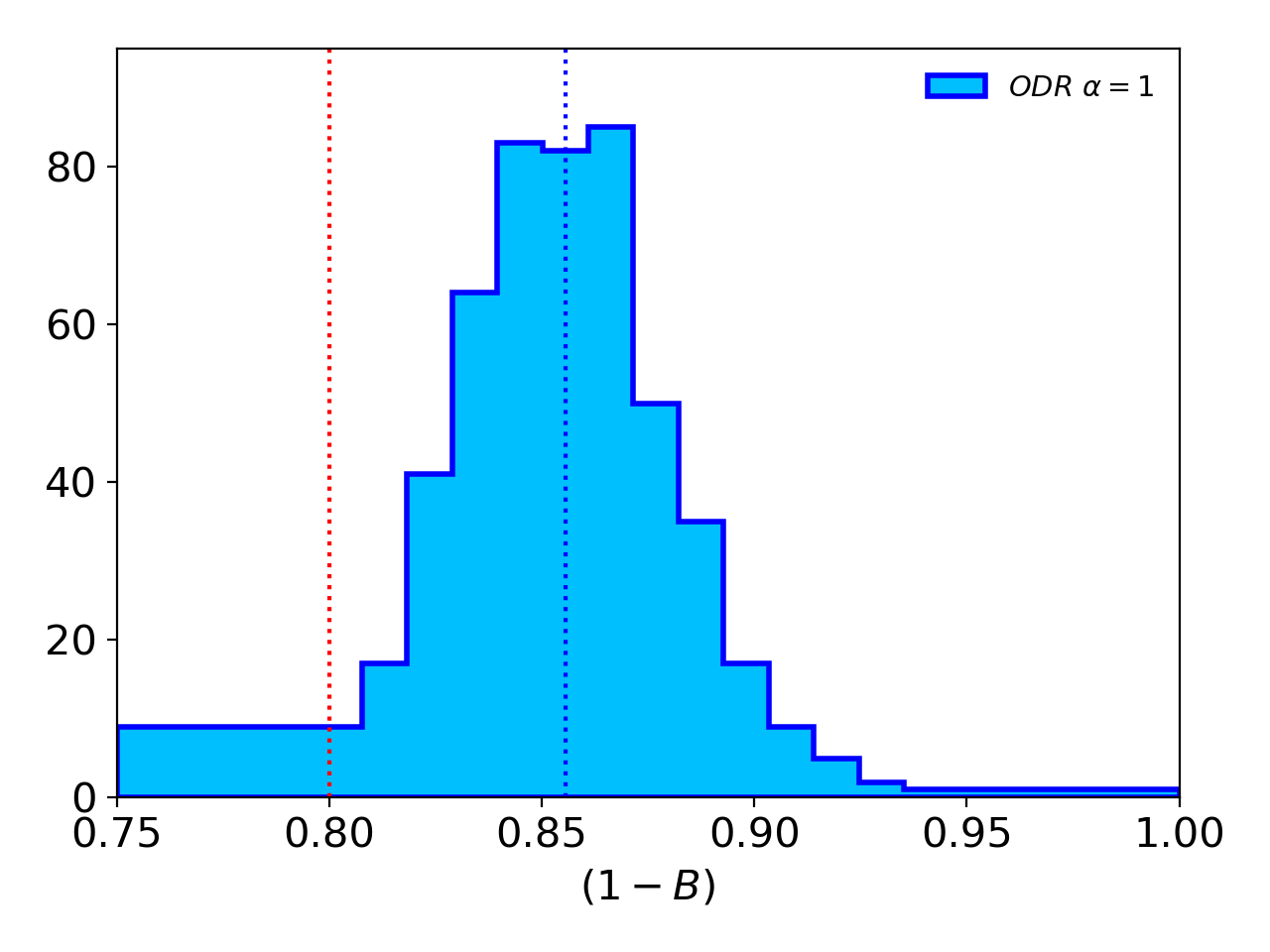}
        \caption{Distribution of the mass bias parameter estimates using the ODR method to fit Eq.~\ref{eq:fit_log} with the slope fixed to 1. The vertical dotted lines represent the mean (recovered) value and the simulated (input) value $(1-B)=0.8$ in blue and red, respectively.}
        \label{fig:odr_cor}
\end{figure}

Figure~\ref{fig:odr_cor} shows that after 500 realisations with realistic noise levels, the ODR method retrieves a mean mass bias parameter that is $\sim 6\,\%$ larger than the input bias, $(1-B) = 0.8$, for the case of a fixed slope ($\alpha=1$). We also obtain the same bias of $\sim 6\,\%$ when the slope was left as a free parameter. 
This value was used later to correct for the final results.

\begin{figure*}
\begin{center}
\includegraphics[trim=0.4cm 0.4cm 0.4cm 0.4cm, clip, width=0.49\textwidth]{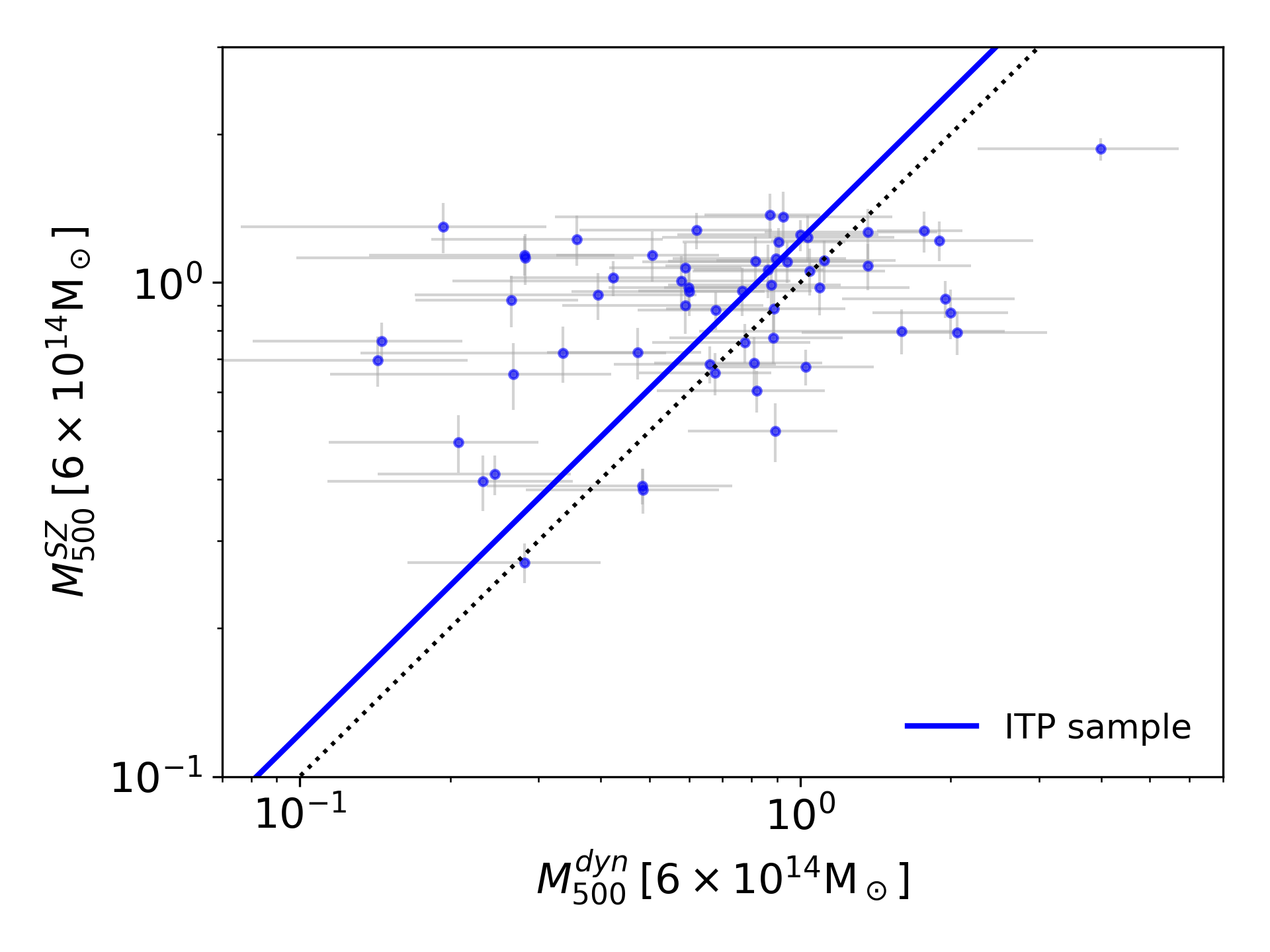}
\includegraphics[trim=0.4cm 0.4cm 0.4cm 0.4cm, clip, width=0.49\textwidth]{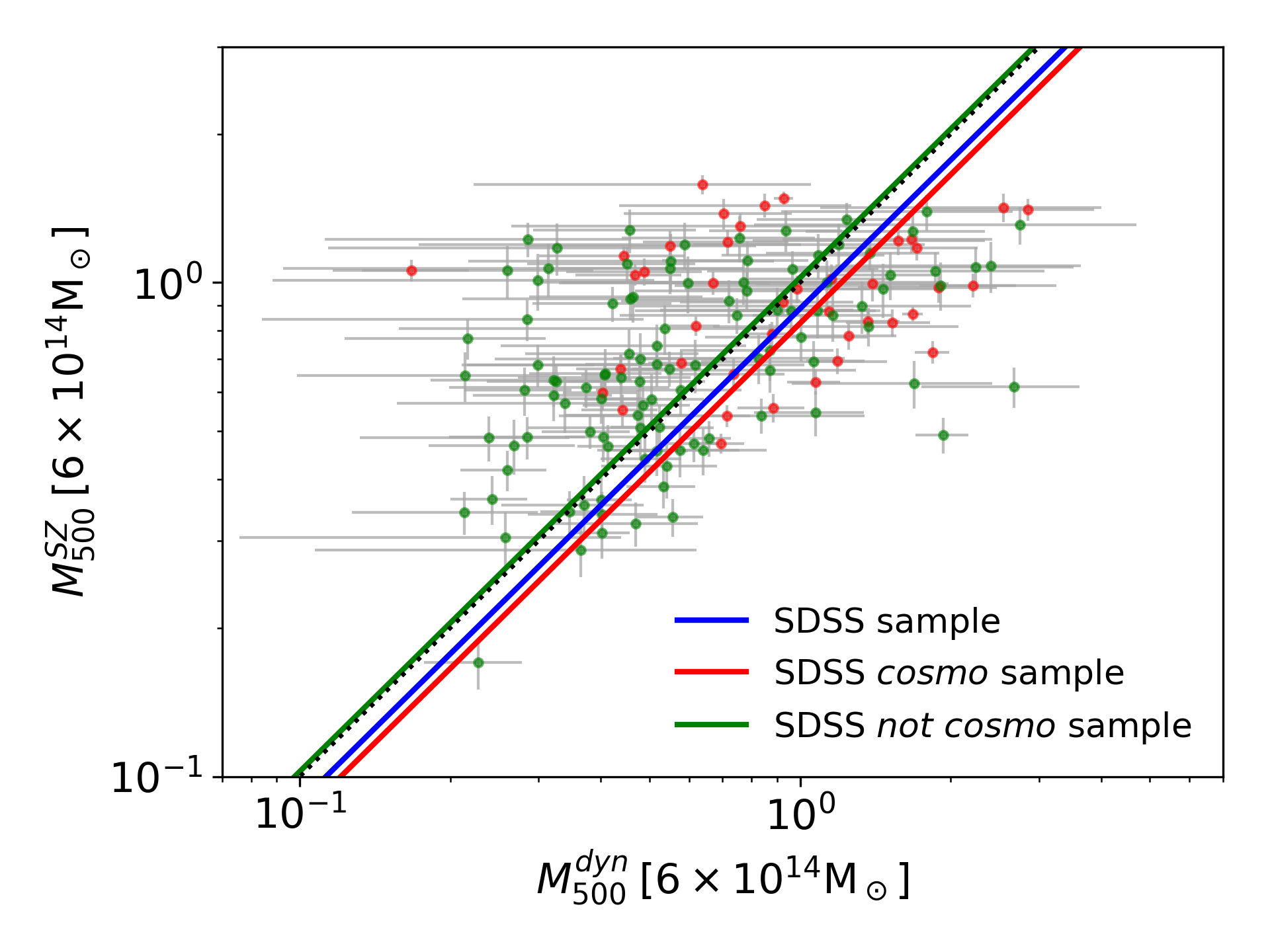}
\end{center}
\caption{$M_{500}^{\rm dyn}-M_{500}^{\rm SZ}$ scaling relation (Eq.~\ref{eq:fit_log} with the slope fixed to 1) for the ITP (left panel) and SDSS (right panel) samples. Red and green dots represent the SDSS clusters included in and excluded from the PlCS, respectively. Red, green, and blue lines are the best fit for the SDSS \textit{cosmo}, the SDSS \textit{not cosmo,} and the whole SDSS sample, respectively, whereas the dotted line is the $1:1$ relation. The ITP and the SDSS fits are at $2.6-\sigma$. his difference between these fits suggests a possible Eddington bias contamination.x}
 \label{fig:relation_ITP_SDSS}
\end{figure*}

\subsection{Mass bias for the ITP and SDSS sub-samples}
Figure \ref{fig:relation_ITP_SDSS} shows our fit for the ITP sample (left panel) and the SDSS sample (blue line in the right panel). Apparently, we find a different mass bias for the two sub-samples,
\begin{align} \label{eq:(1-B)_ed_b}
\left( 1-B \right)_{ITP\phantom{S}} = 1.22 \pm 0.10;\\
\left( 1-B \right)_{SDSS} = 0.89 \pm 0.08.
\end{align}
Because the clusters in the two samples were analysed in the same way, this discrepancy cannot be explained by a heterogeneous mass calculation. 
We ascribe this discrepancy to the different $S/N$ distribution of cluster samples. The median $S/N$ for the ITP sample is $4.9$, whereas the SDSS sample median S/N is $5.7$. Moreover, if we divide the SDSS sample into two parts, the \textit{cosmo} (the clusters in the PlCS) and the \textit{not cosmo},  we find that they have a median $S/N$ $9.4$ and $5.2$, respectively. The fit performed with these two new sub-samples, shown in the right panel of Figure~\ref{fig:relation_ITP_SDSS}, yields $\left( 1-B \right) = 0.83 \pm 0.11$ (red line) and $\left( 1-B \right) = 1.02 \pm 0.08$ (green line) for the \textit{cosmo} and \textit{not cosmo}, respectively. 
\begin{figure}
\begin{center}
\includegraphics[trim=0.4cm 0.4cm 0.4cm 0.4cm, clip, width=0.49\textwidth]{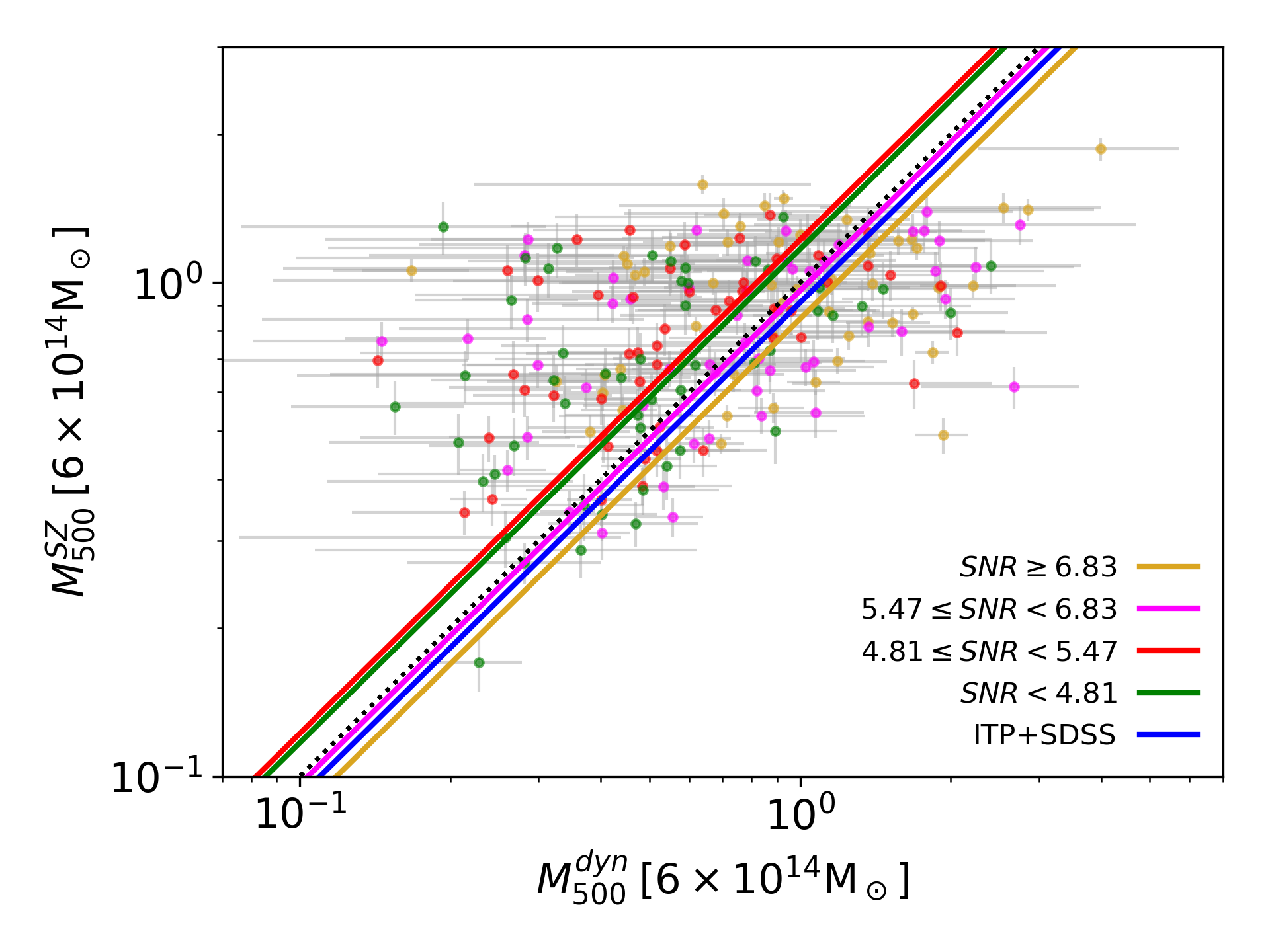}
\end{center}
\caption{Fit of the scaling relation $M_{500}^{\rm dyn}-M_{500}^{\rm SZ}$ in Eq.~\ref{eq:fit_log} (with the slope fixed to 1) for the whole sample ITP+SDSS (blue line). Green, red, magenta, and yellow dots and lines represents the clusters within the four $S/N$ bins: $S/N < 4.81$, $4.81\leq $S/N$ < 5.47$, $5.47\leq $S/N$ < 6.83,$ and $S/N \geq 6.83$ and their best fits, respectively. These fits are only marginally compatible within a maximum difference of $2.9-\sigma$. The dotted line represents the 1:1 relation.}
 \label{fig:sc_eb_SNR}
\end{figure}

This evidence led us to investigate the effect of the $S/N$ of clusters on the $(1-B)$ fits in more detail. 
We divided the whole catalogue (ITP+SDSS) into four $S/N$ bins with the same number of clusters, performing the $M_{\rm dyn}-M_{\rm SZ}$ fit in each bin and for the whole sample. Figure~\ref{fig:sc_eb_SNR} shows that the mass bias decreases from lower to higher value of $S/N$ (see the second column of Table~\ref{table:SNR_bebc}). This difference suggests that our clusters might be affected (totally or partially) by Eddington bias \citep[hereafter EB,][]{eddington1913}.
The statistical noise scatters above the $S/N$ threshold some objects with a mass that is lower than the observational limit, resulting in an over-estimation of these masses.

\begin{table}
\small
\caption{Value of the mass bias before and after the Eddington bias correction for the S/N bins.}
\label{table:SNR_bebc}
\centering
\begin{tabular}{ccc}
\noalign{\smallskip}
\hline\hline
\noalign{\smallskip}
S/N bin  & \multicolumn{2}{c}{$(1-B)$} \\
\cline{2-3}
\noalign{\smallskip}
  & {EB not corrected} & {EB corrected} \\
\hline
\noalign{\smallskip}
$ \qquad \quad S/N < 4.82$     & $1.17 \pm 0.09$ & $0.99 \pm 0.07$ \\
$4.82\leq S/N < 5.47$              & $1.22 \pm 0.07$ & $1.08 \pm 0.06$ \\
$5.47\leq S/N < 6.84$              & $0.96 \pm 0.07$ & $0.89 \pm 0.06$ \\
$\; \qquad \quad S/N \geq 6.84$ & $0.84 \pm 0.11$ & $0.84 \pm 0.11$ \\

\noalign{\smallskip}
\hline
\hline
\end{tabular}
\end{table}

We corrected for this EB effect by using the curves in \cite{remco16}. These were obtained by constructing a $30.000\,$deg$^2$ mock SZ catalogue that took the \cite{tinker08} halo mass function, the redshift-dependent comoving volume element, and the \planck{} noise maps properties into account. They reproduced the effect of the noise on the \emph{\textup{true}} $S/N$ as a random Gaussian variable added to the \emph{\textup{true}} $S/N$. They showed that the effect of the bias is more severe the lower the threshold, and it becomes extremely strong for $S/N \leq 4.5$. Moreover, they found that the effect of the EB also depends on the redshift. It affected the high-redshift clusters more because the halo mass function at a given S/N is steeper.
Although these curves were created using the PSZ2 noise maps, they are a good approximation for the PSZ1 as well because they only depend on the mean S/N level and not on the particular noise level on the PSZ2 maps.

\begin{figure}
\begin{center}
\includegraphics[trim=0.4cm 0.4cm 0.4cm 0.4cm, clip, width=0.49\textwidth]{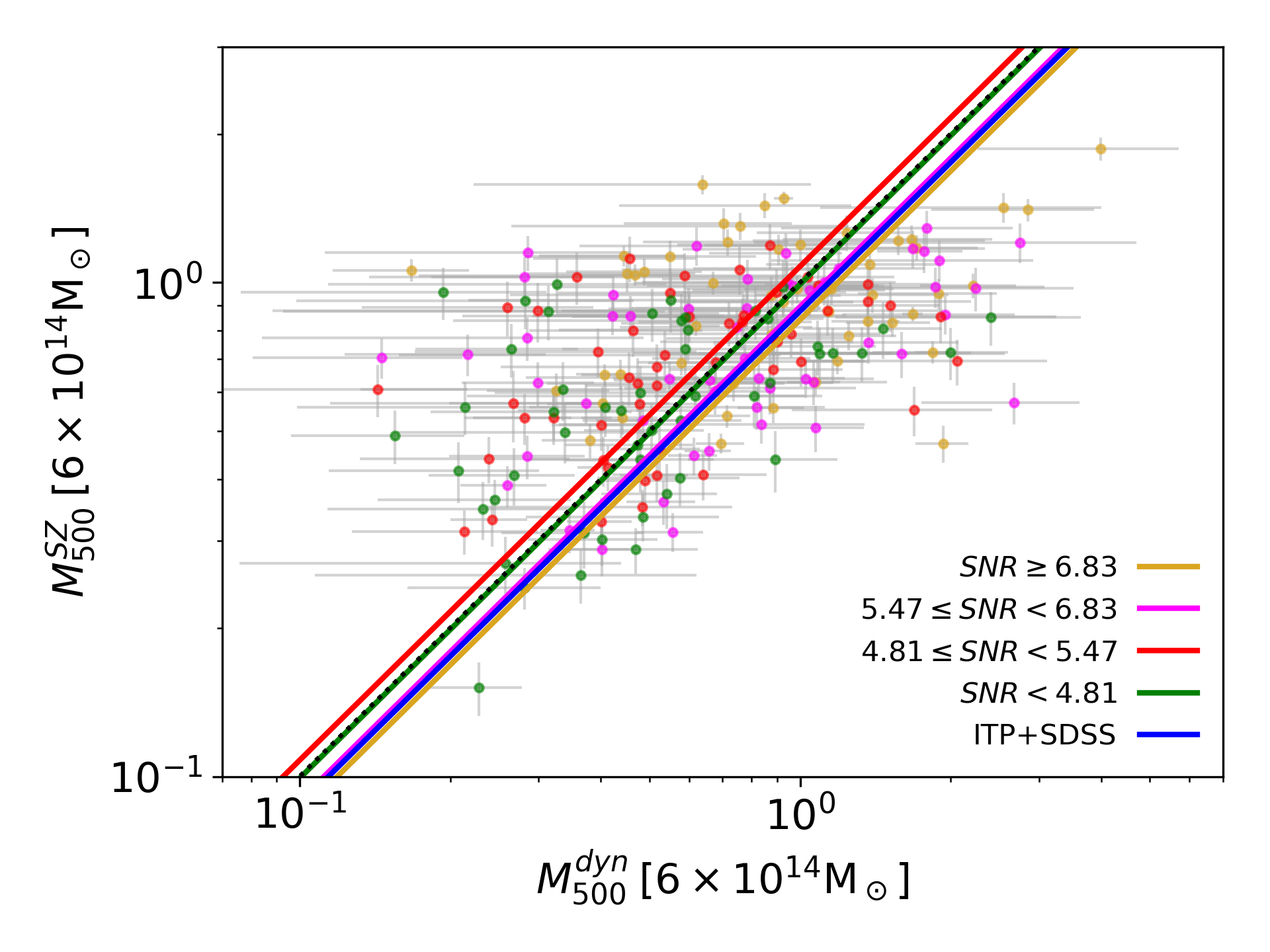}
\end{center}
\caption{Scaling relation $M_{500}^{\rm dyn}-M_{500}^{\rm SZ}$ (fit of Eq.~\ref{eq:fit_log} with a fixed slope equal to 1) after correction for Eddington bias. Magenta, red, yellow, and green dots represent the clusters within the four $S/N$ bins: $S/N < 4.82$, $4.82\leq $S/N$ < 5.47$, $5.47\leq $S/N$ < 6.84,$ and $S/N \geq 6.84$, respectively. After correcting for the Eddington bias, performed following the prescriptions by \cite{remco16}, the best fits within the $S/N$ bins (green, red, magenta, and yellow lines) are now compatible within $1.9-\sigma$ at most with each other and with respect to the best fit of our whole sample (blue line). The dotted line represents the 1:1 relation. }
\label{fig:sc_eb_cor}
\end{figure}

Figure~\ref{fig:sc_eb_cor} shows the $M_{500}^{\rm dyn}-M_{500}^{\rm SZ}$ fit for the whole sample (blue line) and for the $S/N$ bins (green, yellow, red, and magenta circles and lines), and after the EB correction, all the $\left( 1-B \right)$ values are clearly statistically compatible with each other within the $2-\sigma$ level. The corrected values are listed in the last column of Table~\ref{table:SNR_bebc}.

As shown in \cite{remco16}, the magnitude of the EB depends on redshift. We quantified this by analysing the behaviour of the mass bias parameter as a function of redshift. Table~\ref{table:z_bebc} shows that the uncorrected SZ masses lead to different values of $\left( 1-B \right)$. In particular,  the highest redshift bin differs at almost $3-\sigma$ from the lowest bin. 
On the other hand, the $M_{500}^{\rm dyn}$--$M_{500}^{\rm SZ}$ relations after the EB correction show more comparable values of the $\left( 1-B \right)$: all redshift bins are consistent within the $2-\sigma$ level. 
Here we note that the bin containing the clusters at the redshift interval $[0.11$, $0.19]$ yields the highest $(1-B)$ value. 

After the EB correction, the mass bias estimated for the two samples became
\begin{align} \label{eq:(1-B)_ed_b_cor}
\left( 1-B \right)_{ITP\phantom{S}} = 1.06 \pm 0.08;\\
\left( 1-B \right)_{SDSS} = 0.87 \pm 0.08,
\end{align}
which is compatible within the $2-\sigma$ level.

We also investigated the possible dependence of the estimated mass bias on the number of galaxies $\ngal$ used to extract the dynamical masses after correction for EB. 
To do this, each sub-sample (ITP and SDSS) was sub-divided into two bins with the same number of s,  according to the $\ngal$ values. We labelled the bin containing lower (LN) and higher (HN) values $\ngal$ than the median. We repeated our analysis for each of these bins. For the ITP sample, we obtained\begin{align} \label{eq:(1-B)_ITP_ngal}
\left( 1-B \right)_{ITP\phantom{S}}^{LN} = 1.38 \pm 0.23 \\ \left( 1-B \right)_{ITP\phantom{S}}^{HN} = 0.98 \pm 0.08,
\end{align}
whereas the fit for the SDSS sample gave
\begin{align} \label{eq:(1-B)_SDSS_ngal}
\left( 1-B \right)_{SDSS}^{LN} = 1.06 \pm 0.07 \\
\left( 1-B \right)_{SDSS}^{HN} = 0.85 \pm 0.08.
\end{align}
The bins for LN and HN are compatible within the $2-\sigma$ level for both ITP and SDSS samples. However, when we consider the whole sample, the results for the LN and HN sub-samples differ slightly (at the $2.3-\sigma$ level),
\begin{align} \label{eq:(1-B)_ngal}
(1-B)^{LN} = 1.11 \pm 0.07 \\
(1-B)^{HN} = 0.86 \pm 0.08.
\end{align}
Although the residual effects due to $\ngal$ should be accounted for in our method \citep{Ferragamo20}, this result suggests that there might be some marginal residual dependence on $\ngal$ in our velocity and mass estimators.

\begin{table}
\small
\caption{Value of the mass bias before and after correction for Eddigton bias for the redshift bins. }
\label{table:z_bebc}
\centering
\begin{tabular}{ccc}
\noalign{\smallskip}
\hline\hline
\noalign{\smallskip}
redshift bin  & \multicolumn{2}{c}{$(1-B)$} \\
\cline{2-3}
\noalign{\smallskip}
  & {EB not corrected} & {EB corrected} \\
\hline
\noalign{\smallskip}
$  \qquad \quad z < 0.11$     & $0.84 \pm 0.10$ & $0.82 \pm 0.09$ \\
$0.11\leq z < 0.19$               & $1.17 \pm 0.12$ & $1.10 \pm 0.11$ \\
$0.19 \leq z < 0.34$               & $1.05 \pm 0.08$ & $0.96 \pm 0.08$ \\
$\; \qquad \quad z \geq 0.34$  & $1.19 \pm 0.07$ & $1.06 \pm 0.06$ \\
\noalign{\smallskip}
\hline
\hline
\end{tabular}
\end{table}

\subsection{Final result for the mass bias}
\label{subsec:mass_bias_our}

After the EB correction, by fitting the $M_{500}^{\rm dyn}$--$M_{500}^{\rm SZ}$ relation for the whole GCs sample and fixing the slope to unity, we obtain
\begin{equation}
\left( 1-B \right) = 0.88 \pm 0.07.
\label{eq:(1-B)_final1}
\end{equation}
Each of the parameters and related errors shown so far was estimated as the mean and standard deviation of 10000 bootsrap resamples, respectively.
However, taking the expected $\sim 6\%$ bias introduced by the linear regression estimator into account, the corrected value of the mass bias should be
\begin{equation}
\left( 1-B \right) = 0.83 \pm 0.07 \ (stat) \pm 0.02 \ (syst),
\label{eq:(1-B)_final}
\end{equation}
where the first error is the statistical error, and the second error is a systematic error associated with the bias of the ODR method.

The \planck{} analysis showed that the mass bias could be a function of the cluster mass \citep{planck13_count}. For this reason, we repeated our fit with the ODR method, but letting the slope free as well. In this case, after correcting for the bias due to the ODR method, we obtain
\begin{equation}
\frac{M_{500}^{\rm SZ}}{6\times10^{14}\;\msun} = \left(0.84\pm0.12\pm0.02 \right)
\left(\frac{M_{500}^{\rm dyn}}{6\times10^{14}\;\msun} \right)^{1.00\pm0.23\pm0.08}
\label{eq:(1-B)_final_slope_pivot222}
\end{equation}
at the pivot mass $6\times 10^{14}\msun$. This result, which is compatible within $1-\sigma$ with the result in Eq. \ref{eq:(1-B)_final}, does not give indications about a mass dependence for the $M^{\rm dyn}$--$M^{\rm SZ}$ relation.
\begin{figure*}
\begin{center}
\includegraphics[trim=0.4cm 0.4cm 0.2cm 0.4cm, clip, width=0.8\textwidth]{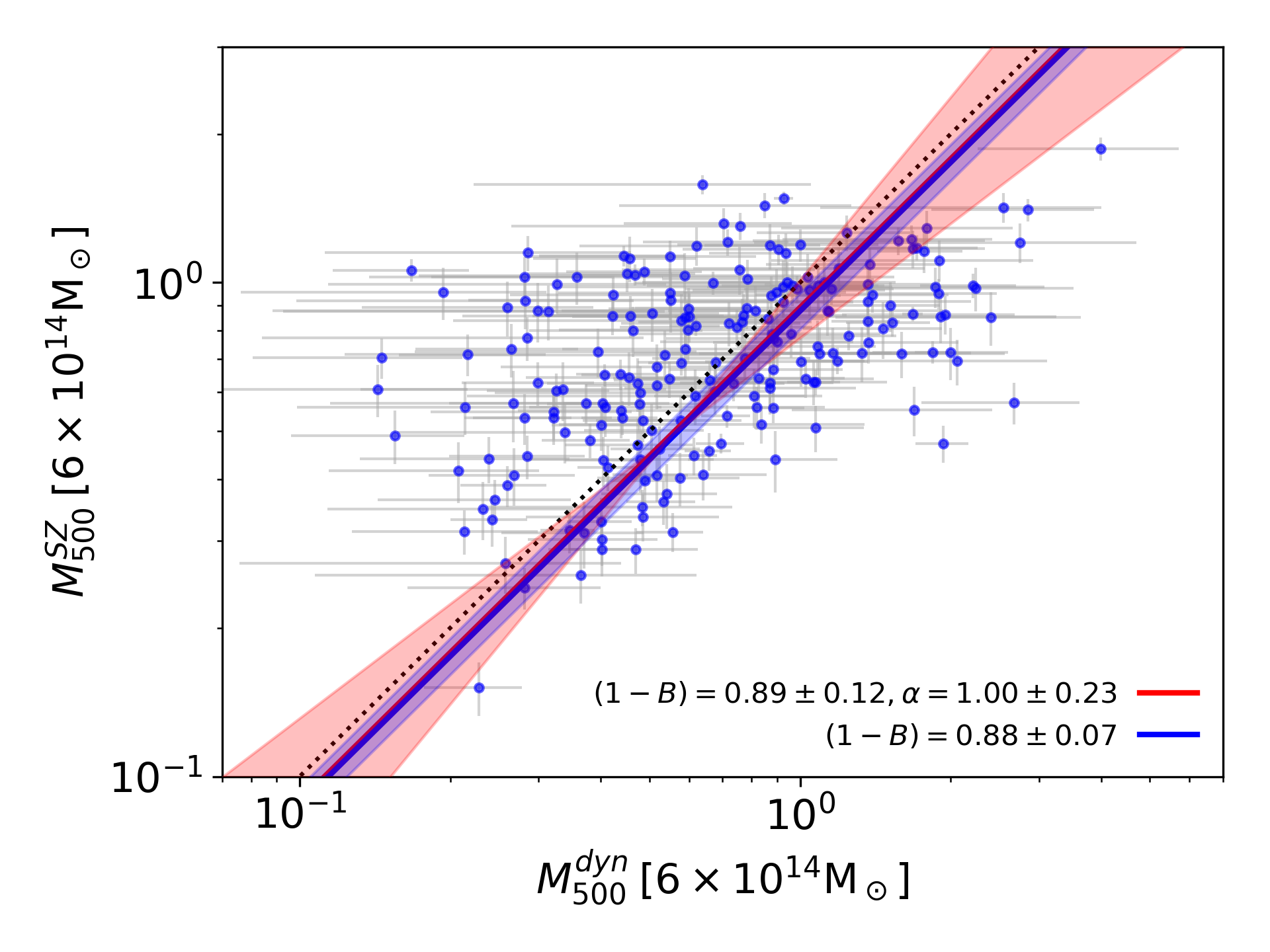}
\end{center}
\caption{Scaling relation between $M_{500}^{\rm SZ}$ and $M_{500}^{\rm dyn}$ for the combined whole sample. The blue line represents the relation obtained by fixing the slope to $\alpha = 1$ (blue line), and the red line represents the relation obtained when the slope was free to vary. The correction of the bias introduced by the ODR estimator is not applied here.}
 \label{fig:scal_rel_final}
\end{figure*}
Figure \ref{fig:scal_rel_final} shows the two fits of the $M_{500}^{\rm dyn}$--$M_{500}^{\rm SZ}$ scaling relation, with the slope fixed to 1 (blue line) and with the slope free to vary (red line).

\section{Comparison with other works}
\label{sec:comparison}
\begin{figure*}
\begin{center}
\includegraphics[trim=0.4cm 0.35cm 0.35cm 0.35cm, clip, width=
0.8 \textwidth]{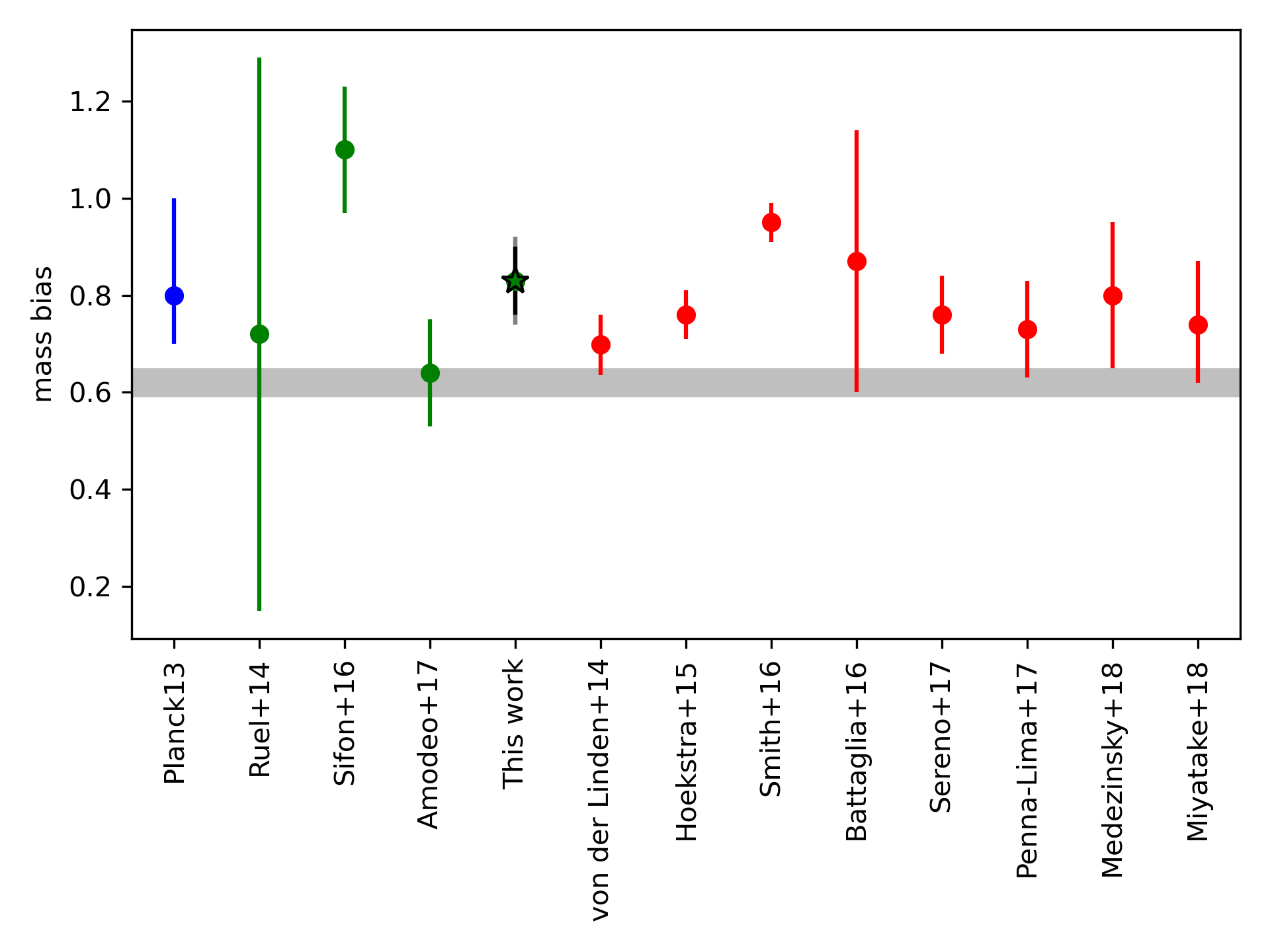}
\end{center}
\caption{Value of the mass bias from previous studies. In blue we show the result from \cite{planck13_count}, using a scaling relation from X-ray observations; in green we plot the mass bias from $M_{\rm dyn}$-$M_{\rm SZ}$ scaling relations, and in red we show those from weak-lensing studies. All these values are listed in table \ref{table:(1-b)_literature}. The grey shaded region represents the mass bias values that reconciles the tension between CMB and SZ number counts from \cite{planck13_count}. The green star represents the mass bias value we found here.}
 \label{fig:comp_literarure_final}
\end{figure*}

The multi-component nature of GCs and multi-wavelength studies allows us to assess the relative bias of mass proxies based on different observables. Here we compare our results on $(1-B)$ with those obtained by other authors using different mass proxies and methods. The literature about this argument is very rich. We restrict our comparison to some of the recent works.  Table~\ref{table:(1-b)_literature} and Figure~\ref{fig:comp_literarure_final} summarise all values obtained by all the surveys described below. 

\subsection{Mass bias from \planck\ Collaboration analysis}
\label{subsec:planck_bias}

The \planck{} Collaboration calibrated the PSZ1 masses using X-ray observations of nearby relaxed clusters. For this reason, the mass bias they found is homologous to the hydrostatic equilibrium mass bias, $M_{500}^{\rm HE}$. 
To find the value of the mass bias $M_{500}^{\rm HE}=(1-b)M_{500}^{\emph{true}}$, they compared the $Y_{500}$--$M_{500}$ relation derived using data from seven  simulations including different types of input physics \citep{nagai06, yang10, sehgal10, krause12, battaglia12, kay12, sembolini13}, with the relation obtained by comparing $Y_{500}$ and the hydrostatic mass \citep{planck13_count}.
The main conclusion is that the dependence of bias on mass,  $b\left(M_{500}^{\emph{true}}\right)$, is strong, which also translates into different slopes for the observed and true $Y_{500}$--$M_{500}$ relations. 
Because of this mass dependence, the \planck{} collaboration selected a representative value for $(1-b)$ as the median value obtained for a mass pivot point $M_{500}=6 \times 10^{14}\, \rm{M_\odot}$ \citep{planck13_count},
\begin{equation}
\left(1-b\right) = 0.8_{-0.1}^{+0.2}.
\label{eq:(1-b)_planck}
\end{equation}
The reported confidence interval $[0.7, 1]$ accounts for the scatter of the different simulations and measurements. 

\subsection{Bias from dynamical mass estimates}
\label{subsec:dynamical_bias}

Several groups have studied the relation $M^{\rm dyn}$--$M^{\rm SZ}$ using SZ data from the \planck{} PSZ2 catalogue \citep{amodeo17}, the Atacama Cosmology Telescope (ACT) \citep{sifon16}, the South Pole Telescope (SPT) \citep{ruel14}, and spectroscopic data from the Gemini Multi-Object Spectrograph (GMOS), installed at the Gemini telescopes. 
Although these  studies also compared the SZ and dynamical masses, each shows important differences in the method with respect to our analysis. 

\subsubsection{South Pole Telescope cluster sample}
\label{subsubsec:spt_bias}
The first of these analyses was published by \cite{ruel14}. Their analysis used 43 SZ-selected GCs within the SPT catalogues \citep{Vanderlinde10, Williamson11, Reichardt13}. The sample consists of massive ($2.7 \leq M_{500}^{\rm SPT} / 10 ^{14}\; \msun  \leq 18.0)$ s at $z \geq 0.3$ with more than 15 members.
The most important differences with respect to our study are the scaling relation between observables and masses. From the SZ point of view, $M_{500}^{\rm SPT}$ were calculated using the $Y$--$M$ scaling in \cite{Reichardt13}. This relation presents a slightly fainter slope than the slope from \cite{planck13_count}. This might result in some differences in the mass estimate, especially for the most massive clusters. From the dynamical mass point of view, the SPT group used the \cite{saro13} relation, which, as shown in \cite{sifon16}, tends to overestimate the dynamical masses (especially for massive clusters) when compared to the \cite{munari13} relation we used here.
Another difference is that following the prescription in \cite{beers90}, they used the biweight as the velocity dispersion estimator applied on cluster members selected by an iterative $3-\sigma$ clipping around the mean velocity. 
In contrast to what we did here, Ruel and collaborators decided to scale the SZ masses from $M_{500}^{\rm SPT}$ to $M_{200}^{\rm SPT}$. However, in order to perform this transformation, they also assumed the same NFW density profile and the \cite{duffy08} mass-concentration relation as in our method. Finally, \cite{ruel14} found that 
\begin{equation}
\exp\Bigg\{ \left\langle \ln\left(\frac{M_{200}^{\rm SPT}}{M_{200}^{\rm dyn}} \right) \right\rangle \Bigg\} = \left( 1-B \right) = 0.72 \pm 0.57,
\label{eq:(1-b)_ruel_SPT}
\end{equation}
which is compatible with our result, except for its large error.

We would like to remark that we report the mass bias \citep[Eq.~16 in][]{ruel14} in Eq.~\ref{eq:(1-b)_ruel_SPT}, according to our definition of $(1-B)$ and after propagating the uncertainties.

\subsubsection{Atacama Cosmology Telescope cluster sample}
\label{subsubsec:act_bias}

The second scaling relation, performed using dynamical masses, was obtained by \cite{sifon16}.
For their cosmological analysis, they used a subsample of ACT clusters composed of 21 objects detected with an $S/N > 5.1$ at redshifts $0.2 \leq z \leq 1.06$ in a mass interval $1.0 < M_{200}^{\rm dyn} < 13.0$ (in units of $10^{14}$\,$\msun$), which constitute their \textit{\textup{cosmological sample}}. The 21 GCs were observed spectroscopically using two different spectrographs: GMOS \citep{Hook04} at the Gemini-South telescope (Chile), and RSS \citep{Burgh03} at the SALT telescope (South Africa). They obtained a median number of observed members of $\ngal  55$.
In this case, there are fewer differences between our method and the method of the ACT group. They used the same scaling relations $Y_{500}$--$M_{500}$ and $\sigma_{200}$--$M_{200}$, the pressure profile by \cite{Arnaud10}, and the AGN fit by \cite{munari13}. The main difference is on the mass-concentration relation that was used to scale $M_{200}^{\rm dyn}$ to $M_{500}^{\rm dyn}$. While we used the \cite{duffy08} relation, they considered the \cite{dutton14} relation.
In addition to the member selection algorithm, they also used the biweight as a velocity dispersion estimator. It is also important to remark that \cite{sifon16} applied an aperture correction to the velocity dispersion estimate and also took the Eddington bias into account when they corrected the $Y_{500}$ estimation, as explained in \cite{hasselfield13}.
In this case, they did not fit for the relation $M_{\rm dyn}$ to $M_{\rm SZ}$, but they defined the mass bias as the ratio of the mean SZ and the mean dynamical mass. They obtained 
\begin{equation}
\frac{\left \langle M_{\rm SZ} \right \rangle}{\left \langle M_{\rm dyn} \right \rangle}= \frac{\left( 1-b_{\rm SZ} \right)}{\beta_{\rm dyn}}= 1.10 \pm 0.13^{stat} \pm 0.14 ^{syst},
\label{eq:(1-b)_sifon_ACT}
\end{equation}
where $\beta_{\rm dyn}$ is the bias of dynamical masses, defined as $\left \langle M_{\rm dyn} \right \rangle = \beta_{\rm dyn} \left \langle M_{\emph{true}} \right \rangle$.
Our result (Eq. \ref{eq:(1-B)_final}) and that obtained by \cite{sifon16} are compatible within $1.7-\sigma$.

\subsubsection{\planck{} PSZ2 sample}
\label{subsubsec:amodeo_bias}

The most recent published scaling relation was obtained by \cite{amodeo17} using 17 low-redshift ($z<0.5$) clusters from the PSZ2 catalogue in the mass interval $3.0 < M_{200}^{\rm dyn} < 14.0$ (units of $10^{14}\,\msun$) and with a median $\ngal = 20$.
In this case, the SZ masses were estimated following the same prescription as we considered here. However, they selected a different velocity dispersion-mass relation. This choice was dictated by the method they used to determine the bias. Similarly to the \planck{} Collaboration, they decided not to compare the masses themselves, but to do the comparison with the $\sigma_{200}$--$M_{200}$ scaling relation normalisation parameter by assuming a self-similar slope ($\alpha=1/3$). To do this, they rescaled the \planck{} $M_{500}^{\rm SZ}$ to $M_{200}^{\rm SZ}$ by assuming the mass-concentration relation by \cite{dutton14}, and fitted the $\sigma_{200}^{\rm obs}-M_{200}^{\rm SZ}$ relation with a fixed slope $\alpha=1/3$. They obtained the normalisation parameter $A=1158 \pm 61$ in this way. 
They compared this result with the \cite{evrard08} relation, which was constrained with an N-body DM-only simulation with a normalisation parameter $A_{DM}=1082.9 \pm 4.0$, from which they obtained 
\begin{equation}
\left( 1-b \right) = \left( \frac{A_{DM}}{A} \right)^3\ b_v^3\ f_{EB}\ f_{corr} = 0.64 \pm 0.11,
\label{eq:(1-b)_amodeo}
\end{equation}
where $b_v$, $f_{EB}$ , and $f_{corr}$ are the velocity bias, the Eddington bias correction, and the correction for the correlated scatter between velocity dispersion and\planck{} masses, respectively.
The velocity bias is defined as $b_v= A_{gal}/A_{DM}$, which arises from assuming that the DM and the galaxies of a cluster can have different velocities. \cite{munari13} used a hydrodynamical simulation and fitted the $\sigma$--$M$ relation considering both the DM particles and galaxies. They obtained two different values for the normalisation parameters, which leads to a velocity bias $b_v=1.08$. In this paper, we accounted for the velocity bias using the AGN fit by \cite{munari13}. Instead, the \cite{evrard08} fit was made with DM-only simulations. 
The Eddington bias correction in this case is a global correction quantified in $f_{EB}=0.84 \pm 0.027$ and the correlated scatter correction arises from the \cite{stanek10} study on the covariance between observables, which is estimated using the Millennium Gas Simulation \citep{hartley08}. \cite{stanek10} found a significant correlation between velocity dispersion and SZ signal due to non-gravitational processes in GCs. \cite{amodeo17} quantified this bias as $f_{corr}=1.01$. 

Comparing our result ($\left( 1-B \right) = 0.83 \pm 0.07 \pm 0.02$) with equation~\ref{eq:(1-b)_amodeo}, we see that they are consistent at $\sim 1.3-\sigma$.

\subsection{Weak-lensing mass bias }
\label{subsec:WL_bias}

Several research groups have studied the SZ mass bias, assuming the weak-lensing (WL) mass as the reference \emph{\textup{true}} mass. Here we describe seven of them. 

\subsubsection{WtG project}
\label{subsubsec:WtG_bias}

In 2014, the group of the Weighing the Giants (WtG) project studied the relation of the $M_{500}^{\rm SZ}$ and the $M_{500}^{\rm WL}$ using a sample of 38 GCs in common with the PSZ1 catalogue \citep{von_der_Linden14}.
For a complete description of the WL mass estimate and for the GCs catalogue they used to constrain the scaling relation, see \cite{applegate14}. 
\cite{von_der_Linden14} found through a bootstrap realisation of the unweighted mean of the ratio between WL and SZ masses the following value:
\begin{equation}
\left \langle \frac{M_{500}^{\rm SZ}}{M_{500}^{\rm WtG}} \right \rangle = \left( 1-b \right) = 0.698 \pm 0.062.
\label{eq:(1-b)_WtG}
\end{equation}
This result is compatible at $1.25-\sigma$ with our estimate of the mass bias.

A sub-sample of 22 GCs of the WtG catalogue is part of the PSZ1 PlCS. Repeating the analysis with this sub-sample alone, they obtained 
\begin{equation}
\left( 1-b \right) = 0.688 \pm 0.072.
\label{eq:(1-b)_WtG_cosmo}
\end{equation}
This result is slightly smaller than the previous one, but they are perfectly compatible with each other. This value was used by the \planck{} Collaboration in its 2015 cosmological analysis based on the PSZ2 catalogue as one of the priors on the value of the mass bias.
The WtG group also performed the fit by letting the slope of the power law vary freely. By using the Bayesian linear regression method developed by \cite{kelly07}, they found 
\begin{equation}
\frac{M_{500}^{\rm SZ}}{10^{15}\;\msun} = \pl0.699_{-0.060}^{+0.059}\pr \pl \frac{M_{500}^{WtG}}{10^{15}\;\msun}\pr^{0.68^{+0.15}_{-0.11}}
\label{eq:(1-b)_WtG_alpha1}
\end{equation}
and
\begin{equation}
\frac{M_{500}^{\rm SZ}}{10^{15}\;\msun} = \pl0.697_{-0.095}^{+0.077}\pr \pl \frac{M_{500}^{WtG}}{10^{15}\;\msun}\pr^{0.76^{+0.39}_{-0.20}},
\label{eq:(1-b)_WtG_alpha2}
\end{equation}
for the whole and the cosmological sample, respectively. 
Both results are compatible within the errors with our fit in Eq~\ref{eq:(1-B)_final_slope_pivot222}.

\subsubsection{Canadian Cluster Comparison Project}
\label{subsubsec:CCCP_bias}

\cite{hoekstra15} presented the comparison between the WL masses, estimated by the Canadian Cluster Comparison Project (CCCP), and the masses from the PSZ1 catalogue of 37 GCs, 20 of them with $S/N \geq 7$.
Through a linear fit, they found
\begin{align} \label{eq:(1-b)_CCCP}
\left( 1-b \right) = 0.76 \pm 0.05\\
\left( 1-b \right) = 0.78 \pm 0.07
\end{align}
for the whole sample of 37 clusters and for the 20 with $S/N \geq 7$, respectively.
These two results are perfectly compatible with our estimate of $\left( 1-b \right)$.

During the comparison of their results with those of WtG, \cite{hoekstra15} fitted a power-law function to investigate whether the bias depended on the mass in their data as well. Using the entire sample, they found
\begin{equation}
\frac{M_{500}^{\rm SZ}}{10^{15}\;h_{70}^{-1} \  \rm{M_\odot}} = \left(0.76 \pm 0.04 \right) \times \left(\frac{M_{500}^{CCCP}}{10^{15}\;h_{70}^{-1} \  \rm{M_\odot}} \right)^{0.64\pm0.17}.
\label{eq:(1-b)_CCCP_slope}
\end{equation}
The compatibility between the slope obtained by \cite{von_der_Linden14}, \cite{hoekstra15} and this paper supports the conclusion by \cite{von_der_Linden14} that the bias of \planck{} masses might depend on the cluster mass. 

\subsubsection{Local Cluster Substructure Survey}
\label{subsubsec:LoCuSS_bias}

\cite{smith16} analysed the mass bias with a sample of 44 clusters in common with the Local Cluster Substructure Survey (LoCuSS) and the PSZ2 catalogue. They obtained 
\begin{equation}
\left( 1-b \right) = 0.95 \pm 0.04.
\label{eq:(1-b)_LoCuSS}
\end{equation}
The LoCuSS sample consists of clusters at redshift $0.15 \leq z \leq 0.3$ with WL masses, estimated in \cite{okabe16}, between $2.12 \leq M_{200}^{WL} / 10^{14} \  \rm{M_\odot} \leq 12.75$. In this case, the mass bias is defined as the weighted mean of the logarithmic ratio of SZ and WL masses, as
\begin{equation}
\left( 1-b \right) =  \rm{exp} \left[ \frac{\sum_{i=1}^{n}\ \omega_i\ \rm{ln}\left( \frac{M_{SZ,i}}{M_{WL,i}} \right)}{\sum_{i=1}^{n}\ \omega_i} \right], 
\label{eq:(1-b)_LoCuSS_2}
\end{equation}
with weights defined by 
\begin{equation}
\omega_i = \left[ \left( \frac{\delta M_{SZ,i}}{\langle \delta M_{SZ} \rangle}\right)^2 + \left( \frac{\delta M_{WL}}{\langle \delta M_{WL,i}\rangle} \right)^2 \right]^{-1}.
\label{eq:(1-b)_LoCuSS_3}
\end{equation}

This LoCuSS measurement ($0.95\pm 0.04$) is higher than our value of $\left( 1-B \right)$. Because no Eddington bias correction is cited in \cite{smith16} or in \cite{okabe16}, we decided to repeat the fit by limiting our sample to the clusters within the range $0.15 \leq z \leq 0.3$. As a result, we obtained $\left( 1-B \right) = 1.17 \pm 0.11$. However, this result might be biased by the clusters within the $0.11 \leq z < 0.19$ (see Table~\ref{table:z_bebc} and the discussion in Section~\ref{subsec:relation2}).

\subsubsection{Canada-France-Hawaii Telescope Stripe 82 Survey}
\label{subsubsec:CS82_bias}
The sample used by \cite{battaglia16} to determine the mass bias consists of 19 GCs from the ACT equatorial sample \citep{hasselfield13} observed during the Canada-France-Hawaii Telescope Stripe 82 Survey (CS82). They divided their sample into two $S/N$ bins. The first contains nine clusters with $S/N > 5$, $\left< M_{500}^{\rm SZ} \right>=4.7 \pm 1.0 \ (\times 10^{14}\; \msun)$, and the second consists of ten clusters within the range $4< S/N<5$, with $\left< M_{500}^{\rm SZ} \right>=2.7 \pm 1.0 \ (\times 10^{14}\; \msun)$. The mass bias they found, defined as the ratio of the mean SZ and WL masses, is
\begin{align} \label{eq:(1-b)_strip82}
\left( 1-b \right)_{S/N > 5} = 0.87 \pm 0.27,\\
\left( 1-b \right)_{S/N < 5} = 0.82 \pm 0.36.
\end{align}
For a detailed description of the WL mass estimation, see \cite{battaglia16} and the references therein, and for SZ mass estimates, see \cite{hasselfield13}.
It is important to remark that the ACT SZ masses are corrected for Eddington bias as explained in \cite{hasselfield13}. Furthermore, \cite{battaglia16} estimated that by comparing $31$ clusters in common between ACT and \planck{}, the corrected ACT masses are lower by $0.89$ times than those of \planck{}on average. 
These results are clearly compatible with our estimate of the bias.

\subsubsection{PSZ2LenS}
\label{subsubsec:PSZ2LenS_bias}

\cite{sereno17} used 32 clusters from the PSZ2LenS sample with a published SZ signal in \planck{} catalogues to estimate the $\left( 1-b \right)$ parameter. The PSZ2LenS sample consists of 35 GCs detected by \planck{} and within the sky coverage of the Canada France Hawaii Telescope Lensing Survey \citep[CFHTLenS,][] {heymans12} and the Red Cluster Sequence Lensing Survey \citep[RCSLenS,][]{hildebrandt16}. The PSZ2Lens clusters lie in a wide mass range, $M_{500}^{WL}=0.9-14.8 \times 10^{14}\; \msun$.
SZ masses are taken from \planck{} catalogues \citep{planck13_cat, planck15_cat}, and were therefore calculated as in this paper. For a detailed description of the WL analysis, see \cite{sereno17} and references therein. 
The PSZ2LenS research group, after assuming the Eddington bias correction from \cite{battaglia16} and \cite{sereno15}, found
\begin{equation}
\exp \left(\frac{\ln\left \langle M_{\rm SZ} \right \rangle}{\ln\left \langle M_{\rm WL} \right \rangle} \right) = \left( 1-b \right) = 0.76 \pm 0.08.
\label{eq:(1-b)_PSZ2LenS}
\end{equation}
This value is statistically consistent with our result.
They also used a sub-sample of 15 PSZ2LenS clusters within the \planck{} cosmological sample, finding $\left( 1-b \right) = 0.67 \pm 0.09$. This result is lower but still compatible with our result at $1.2-\sigma$. 

\subsubsection{Cluster Lensing And Supernova survey with Hubble}
\label{subsubsec:CLASH_bias}

The WL masses of 21 GCs from the Cluster Lensing And Supernova survey with Hubble \citep[CLASH,][]{postman12}) were used by \cite{penna17} to compare with the respective SZ masses from \planck{} catalogues. They performed a Bayesian analysis to constrain the mass bias $\left( 1-b \right)$. After the correction of $M_{\rm SZ}$ for Eddington bias from \cite{battaglia16}, they found
\begin{equation}
\left( 1-b \right) = 0.73 \pm 0.10,
\label{eq:(1-b)_CLASH}
\end{equation}
which is fully compatible with our result.

\subsubsection{Hyper Suprime-Cam Subaru Strategic Program}
\label{subsubsec:HSC-SSP_bias}

The Hyper Suprime-Cam Subaru Strategic Program \citep[HSC-SSP,][]{aihara18b, aihara18a} group published two studies for which they compared their WL masses with SZ masses from \planck{} PSZ2 and ACTPol (Atacama Cosmology Telescope Polarimeter experiment).
In the first study, \cite{medezinski18} used five clusters from the HSC-SSP first-year catalogue \citep{planck15_cat} in common with the PSZ2 catalogue. They chose the mean $M_{500}^{\rm WL}$, defined as the weighted stack of the WL signal of these five clusters. The mean SZ mass is the weighted mean of the masses retrieved from the PSZ2 corrected for Eddington bias using the prescription given by \cite{battaglia16}. The weight for the WL stack and for the SZ mean is the same and depends on the errors on the galaxy shape measurement for each cluster. The ratio of the mean WL and SZ mass is the bias
\begin{equation}
\frac{\left \langle M_{\rm SZ} \right \rangle}{\left \langle M_{\rm WL} \right \rangle}  = \left( 1-b \right) = 0.80 \pm 0.14,
\label{eq:(1-b)_HSC_planck}
\end{equation}
which is compatible with our result. 

In a second study, \cite{miyatake18} used eight GCs in common with the ACTPol \citep{hilton18} sample. As in the previous analysis, the mean WL mass is the result of the weighted stack of all the clusters taken into account, whereas $\left \langle M_{500}^{\rm SZ} \right \rangle$ is the weighted mean of the corrected SZ masses from the ACTPol catalogue. The mass bias for this sample is
\begin{equation}
\frac{\left \langle M_{SZ} \right \rangle}{\left \langle M_{WL} \right \rangle}  = \left( 1-b \right) = 0.74_{-0.12}^{+0.13}.
\label{eq:(1-b)_HSC_ACTPol}
\end{equation}
In this case, the value of the mass bias is also compatible with the value obtained in our analysis.

\begin{table*}
        \tiny
        \caption{Summary of mass bias measurements from the literature according to each of the particular definitions in section~\ref{sec:comparison}.}
        \label{table:(1-b)_literature}
        \centering
        \begin{tabular}{c c c c c}
        \noalign{\smallskip}
        \hline\hline
        \noalign{\smallskip}
        \, SURVEY & REFERENCE SAMPLE & N. CLUSTERS & Mass bias & reference \\ 
        \noalign{\smallskip}
        \hline
        \hline
        \noalign{\smallskip}
        \multicolumn{5}{c}{X-RAY } \\
        \noalign{\smallskip}
        \hline
        \noalign{\smallskip}
         & \planck{} PSZ1 & 189  & $0.8_{-0.2}^{+0.1} $ & {\cite{planck13_cat}}\\
        \noalign{\smallskip}
        \hline
        \hline
        \noalign{\smallskip}
        \multicolumn{5}{c}{VELOCITY DISPERSION } \\
        \noalign{\smallskip}
        \hline
        \noalign{\smallskip}
         & SPT & 44 & $0.72 \pm 0.57$ & {\cite{ruel14}}  \\
         & ACT & 21 & $1.10 \pm 0.13$ & {\cite{sifon16}} \\
         & \planck{} PSZ2& 17 & $0.64 \pm 0.11$ & {\cite{amodeo17}} \\
         & \planck{} PSZ1 & 207 & $0.83 \pm 0.07 \pm 0.02 $ & {This work} \\
        \noalign{\smallskip}
        \hline
        \hline
        \noalign{\smallskip}
        \multicolumn{5}{c}{WEAK LENSING } \\
        \noalign{\smallskip}
        \hline
        \noalign{\smallskip}
        WtG    & \planck{} PSZ1 & 38 & $0.688 \pm 0.072$ & {\cite{von_der_Linden14}} \\
        CCCP & \planck{} PSZ1 & 37 & $0.76 \pm 0.05$ & {\cite{hoekstra15}} \\
        LoCuSS  & \planck{} PSZ2 & 44   & $0.95 \pm 0.04$ & {\cite{smith16}} \\
        CS82  & ACT & 19 & $0.87 \pm 0.27$ & {\cite{battaglia16}} \\
        PSZ2LenS  & \planck{} PSZ2 & 32 & $0.76 \pm 0.08$ & {\cite{sereno17}} \\
        CLASH  & \planck{} PSZ1 & 21 & $0.73 \pm 0.10$ & {\cite{penna17}} \\
        HSC-SSP  & \planck{} PSZ2 & 5 & $0.80 \pm 0.15$ & {\cite{medezinski18}} \\
        HSC-SSP  & ACTPol & 8 & $0.74_{-012}^{+0.13}$ & {\cite{miyatake18}} \\
        \noalign{\smallskip}
        \hline
        \end{tabular}
\end{table*}

\section{Conclusions}
\label{sec:cosmo}

This is the third (and last) paper in a series describing the results of the ITP13 observational program, dedicated to the characterisation of the PSZ1 sources in the northern sky without known optical counterparts at the time the catalogue was published. Here we presented for the first time the velocity dispersion and mass estimates for 58 clusters in the PSZ1-North sample and for 35 clusters that are not associated with the PSZ1 sample. 

Using SDSS archival data, we also studied 212 clusters with known counterparts, and we also extracted the velocity dispersion and mass estimation, using the same method as for the ITP sample. This paper presents dynamical masses for 270 s within the PSZ1-North sample.

A sub-sample of 207 clusters was used to explore the mass bias between the dynamical mass and the SZ mass estimates.  
Galaxy cluster number counts are extremely sensitive to $\Omega_m$ and $\sigma_8$ through the mass function. However, the cosmological analyses performed by the \planck{} Collaboration result in a tension between the constraints from the primordial CMB power spectrum and those derived from the cluster number counts \citep{planck13_count, planck15_count}. The main cause of this tension was originally ascribed to the value of the mass bias. The \planck{} Collaboration, through the joint analysis of SZ, X-ray data, and simulations, constrained this parameter to the value of $(1-b)=0.8_{-0.1}^{+0.2}$. 
In this paper, we used the largest sub-sample of PSZ1 clusters (207) observed both in microwave and optical wavelengths to date, and we obtained $(1-B)=0.83\pm 0.07 \ (stat) \pm 0.02 \ (syst)$. This measurement presents the lowest statistical error obtained so far using dynamical masses as reference,  and it is statistically compatible with the \planck{} one.

We have also compared our results with those in the literature. We find that although it is slightly higher than average, our measurement is compatible within $1-\sigma$ in most cases, both with those obtained from dynamical mass analyses and those obtained from comparison with weak-lensing masses. 
The final mass bias from this study, as well as the mean value of all other previous results in the literature, is still higher than the value that is required to reconcile the tension between CMB and SZ number counts \citep{planck13_count, planck15_count}.

\begin{acknowledgements}
We thank R. van der Burg and J.-B. Melin for useful comments.
This article is based on observations made with the Gran
Telescopio Canarias operated by the Instituto de Astrofisica de Canarias, the Isaac Newton Telescope, and the William Herschel Telescope operated by the Isaac Newton Group of Telescopes, and the Italian Telescopio Nazionale Galileo operated by the Fundacion Galileo Galilei of the INAF (Istituto Nazionale di Astrofisica). All these facilities are located at the Spanish Roque de los Muchachos Observatory of the Instituto de Astrofisica de Canarias on the island of La Palma. 
Part of this research has been carried out with telescope time awarded by the CCI International Time Programme at the Canary Islands Observatories (programmes ITP13B-15A).
Funding for the Sloan Digital Sky Survey (SDSS) has been provided by the Alfred P. Sloan Foundation, the Participating Institutions, the National Aeronautics and Space Administration, the National Science Foundation, the U.S. Department of Energy, the Japanese Monbukagakusho, and the Max Planck Society. 
This work has been partially funded by the Spanish
Ministry of Science and Innovation (MICINN) under the projects ESP2013-48362-C2-1-P, AYA2014-60438-P and AYA2017-84185-P. %
AS and RB acknowledge financial support from MICINN under the Severo Ochoa Programs SEV-2011-0187 and SEV-2015-0548.
HL is supported by ETAg grants PUT1627 and PRG1006 and by EU through the ERDF CoE grant TK133.
\end{acknowledgements}

\bibliographystyle{aa} 
\bibliography{biblio.bib} 

\begin{thebibliography}{72}
\expandafter\ifx\csname natexlab\endcsname\relax\def\natexlab#1{#1}\fi

\bibitem[{{Aguado-Barahona} {et~al.}(2020){Aguado-Barahona},
  {Rubi{\~n}o-Mart{\'i}n}, {Ferragamo}, {Barrena}, , {Streblyanska}, \&
  {Tramonte}}]{aguado20}
{Aguado-Barahona}, A., {Rubi{\~n}o-Mart{\'i}n}, J.~A., {Ferragamo}, A.,
  {et~al.} 2020, \aap (submitted)

\bibitem[{{Aihara} {et~al.}(2011){Aihara}, {Allende Prieto}, {An}, {Anderson},
  {Aubourg}, {Balbinot}, {Beers}, {Berlind}, {Bickerton}, {Bizyaev}, {Blanton},
  {Bochanski}, {Bolton}, {Bovy}, {Brandt}, {Brinkmann}, {Brown}, {Brownstein},
  {Busca}, {Campbell}, {Carr}, {Chen}, {Chiappini}, {Comparat}, {Connolly},
  {Cortes}, {Croft}, {Cuesta}, {da Costa}, {Davenport}, {Dawson}, {Dhital},
  {Ealet}, {Ebelke}, {Edmondson}, {Eisenstein}, {Escoffier}, {Esposito},
  {Evans}, {Fan}, {Femen{\'{\i}}a Castell{\'a}}, {Font-Ribera}, {Frinchaboy},
  {Ge}, {Gillespie}, {Gilmore}, {Gonz{\'a}lez Hern{\'a}ndez}, {Gott}, {Gould},
  {Grebel}, {Gunn}, {Hamilton}, {Harding}, {Harris}, {Hawley}, {Hearty}, {Ho},
  {Hogg}, {Holtzman}, {Honscheid}, {Inada}, {Ivans}, {Jiang}, {Johnson},
  {Jordan}, {Jordan}, {Kazin}, {Kirkby}, {Klaene}, {Knapp}, {Kneib},
  {Kochanek}, {Koesterke}, {Kollmeier}, {Kron}, {Lampeitl}, {Lang}, {Le Goff},
  {Lee}, {Lin}, {Long}, {Loomis}, {Lucatello}, {Lundgren}, {Lupton}, {Ma},
  {MacDonald}, {Mahadevan}, {Maia}, {Makler}, {Malanushenko}, {Malanushenko},
  {Mandelbaum}, {Maraston}, {Margala}, {Masters}, {McBride}, {McGehee},
  {McGreer}, {M{\'e}nard}, {Miralda-Escud{\'e}}, {Morrison}, {Mullally},
  {Muna}, {Munn}, {Murayama}, {Myers}, {Naugle}, {Neto}, {Nguyen}, {Nichol},
  {O'Connell}, {Ogando}, {Olmstead}, {Oravetz}, {Padmanabhan},
  {Palanque-Delabrouille}, {Pan}, {Pandey}, {P{\^a}ris}, {Percival},
  {Petitjean}, {Pfaffenberger}, {Pforr}, {Phleps}, {Pichon}, {Pieri}, {Prada},
  {Price-Whelan}, {Raddick}, {Ramos}, {Reyl{\'e}}, {Rich}, {Richards}, {Rix},
  {Robin}, {Rocha-Pinto}, {Rockosi}, {Roe}, {Rollinde}, {Ross}, {Ross},
  {Rossetto}, {S{\'a}nchez}, {Sayres}, {Schlegel}, {Schlesinger}, {Schmidt},
  {Schneider}, {Sheldon}, {Shu}, {Simmerer}, {Simmons}, {Sivarani}, {Snedden},
  {Sobeck}, {Steinmetz}, {Strauss}, {Szalay}, {Tanaka}, {Thakar}, {Thomas},
  {Tinker}, {Tofflemire}, {Tojeiro}, {Tremonti}, {Vandenberg}, {Vargas
  Maga{\~n}a}, {Verde}, {Vogt}, {Wake}, {Wang}, {Weaver}, {Weinberg}, {White},
  {White}, {Yanny}, {Yasuda}, {Yeche}, \& {Zehavi}}]{aihara11_sdss}
{Aihara}, H., {Allende Prieto}, C., {An}, D., {et~al.} 2011, \apjs, 193, 29

\bibitem[{{Aihara} {et~al.}(2018{\natexlab{a}}){Aihara}, {Arimoto},
  {Armstrong}, {Arnouts}, {Bahcall}, {Bickerton}, {Bosch}, {Bundy}, {Capak},
  {Chan}, {Chiba}, {Coupon}, {Egami}, {Enoki}, {Finet}, {Fujimori}, {Fujimoto},
  {Furusawa}, {Furusawa}, {Goto}, {Goulding}, {Greco}, {Greene}, {Gunn},
  {Hamana}, {Harikane}, {Hashimoto}, {Hattori}, {Hayashi}, {Hayashi},
  {He{\l}miniak}, {Higuchi}, {Hikage}, {Ho}, {Hsieh}, {Huang}, {Huang},
  {Ikeda}, {Imanishi}, {Inoue}, {Iwasawa}, {Iwata}, {Jaelani}, {Jian},
  {Kamata}, {Karoji}, {Kashikawa}, {Katayama}, {Kawanomoto}, {Kayo}, {Koda},
  {Koike}, {Kojima}, {Komiyama}, {Konno}, {Koshida}, {Koyama}, {Kusakabe},
  {Leauthaud}, {Lee}, {Lin}, {Lin}, {Lupton}, {Mandelbaum}, {Matsuoka},
  {Medezinski}, {Mineo}, {Miyama}, {Miyatake}, {Miyazaki}, {Momose}, {More},
  {More}, {Moritani}, {Moriya}, {Morokuma}, {Mukae}, {Murata}, {Murayama},
  {Nagao}, {Nakata}, {Niida}, {Niikura}, {Nishizawa}, {Obuchi}, {Oguri},
  {Oishi}, {Okabe}, {Okamoto}, {Okura}, {Ono}, {Onodera}, {Onoue}, {Osato},
  {Ouchi}, {Price}, {Pyo}, {Sako}, {Sawicki}, {Shibuya}, {Shimasaku},
  {Shimono}, {Shirasaki}, {Silverman}, {Simet}, {Speagle}, {Spergel},
  {Strauss}, {Sugahara}, {Sugiyama}, {Suto}, {Suyu}, {Suzuki}, {Tait},
  {Takada}, {Takata}, {Tamura}, {Tanaka}, {Tanaka}, {Tanaka}, {Tanaka},
  {Terai}, {Terashima}, {Toba}, {Tominaga}, {Toshikawa}, {Turner}, {Uchida},
  {Uchiyama}, {Umetsu}, {Uraguchi}, {Urata}, {Usuda}, {Utsumi}, {Wang}, {Wang},
  {Wong}, {Yabe}, {Yamada}, {Yamanoi}, {Yasuda}, {Yeh}, {Yonehara}, \&
  {Yuma}}]{aihara18a}
{Aihara}, H., {Arimoto}, N., {Armstrong}, R., {et~al.} 2018{\natexlab{a}},
  Publications of the Astronomical Society of Japan, 70, S4

\bibitem[{{Aihara} {et~al.}(2018{\natexlab{b}}){Aihara}, {Armstrong},
  {Bickerton}, {Bosch}, {Coupon}, {Furusawa}, {Hayashi}, {Ikeda}, {Kamata},
  {Karoji}, {Kawanomoto}, {Koike}, {Komiyama}, {Lang}, {Lupton}, {Mineo},
  {Miyatake}, {Miyazaki}, {Morokuma}, {Obuchi}, {Oishi}, {Okura}, {Price},
  {Takata}, {Tanaka}, {Tanaka}, {Tanaka}, {Uchida}, {Uraguchi}, {Utsumi},
  {Wang}, {Yamada}, {Yamanoi}, {Yasuda}, {Arimoto}, {Chiba}, {Finet},
  {Fujimori}, {Fujimoto}, {Furusawa}, {Goto}, {Goulding}, {Gunn}, {Harikane},
  {Hattori}, {Hayashi}, {He{\l}miniak}, {Higuchi}, {Hikage}, {Ho}, {Hsieh},
  {Huang}, {Huang}, {Imanishi}, {Iwata}, {Jaelani}, {Jian}, {Kashikawa},
  {Katayama}, {Kojima}, {Konno}, {Koshida}, {Kusakabe}, {Leauthaud}, {Lee},
  {Lin}, {Lin}, {Mandelbaum}, {Matsuoka}, {Medezinski}, {Miyama}, {Momose},
  {More}, {More}, {Mukae}, {Murata}, {Murayama}, {Nagao}, {Nakata}, {Niida},
  {Niikura}, {Nishizawa}, {Oguri}, {Okabe}, {Ono}, {Onodera}, {Onoue}, {Ouchi},
  {Pyo}, {Shibuya}, {Shimasaku}, {Simet}, {Speagle}, {Spergel}, {Strauss},
  {Sugahara}, {Sugiyama}, {Suto}, {Suzuki}, {Tait}, {Takada}, {Terai}, {Toba},
  {Turner}, {Uchiyama}, {Umetsu}, {Urata}, {Usuda}, {Yeh}, \&
  {Yuma}}]{aihara18b}
{Aihara}, H., {Armstrong}, R., {Bickerton}, S., {et~al.} 2018{\natexlab{b}},
  Publications of the Astronomical Society of Japan, 70, S8

\bibitem[{{Akritas} \& {Bershady}(1996)}]{akritas96}
{Akritas}, M.~G. \& {Bershady}, M.~A. 1996, \apj, 470, 706

\bibitem[{{Allen} {et~al.}(2011){Allen}, {Evrard}, \& {Mantz}}]{allen11}
{Allen}, S.~W., {Evrard}, A.~E., \& {Mantz}, A.~B. 2011, \araa, 49, 409

\bibitem[{{Amodeo} {et~al.}(2017){Amodeo}, {Mei}, {Stanford}, {Bartlett},
  {Melin}, {Lawrence}, {Chary}, {Shim}, {Marleau}, \& {Stern}}]{amodeo17}
{Amodeo}, S., {Mei}, S., {Stanford}, S.~A., {et~al.} 2017, \apj, 844, 101

\bibitem[{{Applegate} {et~al.}(2014){Applegate}, {von der Linden}, {Kelly},
  {Allen}, {Allen}, {Burchat}, {Burke}, {Ebeling}, {Mantz}, \&
  {Morris}}]{applegate14}
{Applegate}, D.~E., {von der Linden}, A., {Kelly}, P.~L., {et~al.} 2014,
  \mnras, 439, 48

\bibitem[{{Arnaud} {et~al.}(2010){Arnaud}, {Pratt}, {Piffaretti},
  {B{\"o}hringer}, {Croston}, \& {Pointecouteau}}]{Arnaud10}
{Arnaud}, M., {Pratt}, G.~W., {Piffaretti}, R., {et~al.} 2010, \aap, 517, A92

\bibitem[{{Barrena} {et~al.}(2020){Barrena}, {Ferragamo},
  {Rubi{\~n}o-Mart{\'i}n}, {Streblyanska}, {Aguado-Barahona}, {Tramonte},
  {G{\'e}nova-Santos}, {Hempel}, {Lietzen}, {Aghanim}, {B{\"o}hringer}, {Chon},
  {Dahle}, {Douspis}, {Lasenby}, {Mazzotta}, {Melin}, {Pointecouteau}, {Pratt},
  \& {Rossetti}}]{rafa20}
{Barrena}, R., {Ferragamo}, A., {Rubi{\~n}o-Mart{\'i}n}, J.~A., {et~al.} 2020,
  \aap (accepted for publication), arXiv:2004.07913

\bibitem[{{Barrena} {et~al.}(2018){Barrena}, {Streblyanska}, {Ferragamo},
  {Rubi{\~n}o-Mart{\'\i}n}, {Aguado-Barahona}, {Tramonte}, {G{\'e}nova-Santos},
  {Hempel}, {Lietzen}, {Aghanim}, {Arnaud}, {B{\"o}hringer}, {Chon},
  {Democles}, {Dahle}, {Douspis}, {Lasenby}, {Mazzotta}, {Melin},
  {Pointecouteau}, {Pratt}, {Rossetti}, \& {van der Burg}}]{rafa18}
{Barrena}, R., {Streblyanska}, A., {Ferragamo}, A., {et~al.} 2018, \aap, 616,
  A42

\bibitem[{{Battaglia} {et~al.}(2012){Battaglia}, {Bond}, {Pfrommer}, \&
  {Sievers}}]{battaglia12}
{Battaglia}, N., {Bond}, J.~R., {Pfrommer}, C., \& {Sievers}, J.~L. 2012, \apj,
  758, 74

\bibitem[{{Battaglia} {et~al.}(2016){Battaglia}, {Leauthaud}, {Miyatake},
  {Hasselfield}, {Gralla}, {Allison}, {Bond}, {Calabrese}, {Crichton},
  {Devlin}, {Dunkley}, {D{\"u}nner}, {Erben}, {Ferrara}, {Halpern}, {Hilton},
  {Hill}, {Hincks}, {Hlo{\v{z}}ek}, {Huffenberger}, {Hughes}, {Kneib},
  {Kosowsky}, {Makler}, {Marriage}, {Menanteau}, {Miller}, {Moodley}, {Moraes},
  {Niemack}, {Page}, {Shan}, {Sehgal}, {Sherwin}, {Sievers}, {Sif{\'o}n},
  {Spergel}, {Staggs}, {Taylor}, {Thornton}, {van Waerbeke}, \&
  {Wollack}}]{battaglia16}
{Battaglia}, N., {Leauthaud}, A., {Miyatake}, H., {et~al.} 2016, Journal of
  Cosmology and Astro-Particle Physics, 2016, 013

\bibitem[{{Beers} {et~al.}(1990){Beers}, {Flynn}, \& {Gebhardt}}]{beers90}
{Beers}, T.~C., {Flynn}, K., \& {Gebhardt}, K. 1990, \aj, 100, 32

\bibitem[{{Burgh} {et~al.}(2003){Burgh}, {Nordsieck}, {Kobulnicky}, {Williams},
  {O'Donoghue}, {Smith}, \& {Percival}}]{Burgh03}
{Burgh}, E.~B., {Nordsieck}, K.~H., {Kobulnicky}, H.~A., {et~al.} 2003, in
  Society of Photo-Optical Instrumentation Engineers (SPIE) Conference Series,
  Vol. 4841, Instrument Design and Performance for Optical/Infrared
  Ground-based Telescopes, ed. M.~{Iye} \& A.~F.~M. {Moorwood}, 1463--1471

\bibitem[{{da Silva} {et~al.}(2004){da Silva}, {Kay}, {Liddle}, \&
  {Thomas}}]{daSilva2004}
{da Silva}, A.~C., {Kay}, S.~T., {Liddle}, A.~R., \& {Thomas}, P.~A. 2004,
  \mnras, 348, 1401

\bibitem[{{Duffy} {et~al.}(2008){Duffy}, {Schaye}, {Kay}, \& {Dalla
  Vecchia}}]{duffy08}
{Duffy}, A.~R., {Schaye}, J., {Kay}, S.~T., \& {Dalla Vecchia}, C. 2008,
  \mnras, 390, L64

\bibitem[{{Dutton} \& {Macci{\`o}}(2014)}]{dutton14}
{Dutton}, A.~A. \& {Macci{\`o}}, A.~V. 2014, \mnras, 441, 3359

\bibitem[{{Eddington}(1913)}]{eddington1913}
{Eddington}, A.~S. 1913, \mnras, 73, 359

\bibitem[{{Evrard} {et~al.}(2008){Evrard}, {Bialek}, {Busha}, {White}, {Habib},
  {Heitmann}, {Warren}, {Rasia}, {Tormen}, {Moscardini}, {Power}, {Jenkins},
  {Gao}, {Frenk}, {Springel}, {White}, \& {Diemand}}]{evrard08}
{Evrard}, A.~E., {Bialek}, J., {Busha}, M., {et~al.} 2008, \apj, 672, 122

\bibitem[{{Ferragamo} {et~al.}(2020){Ferragamo}, {Rubi{\~n}o-Mart{\'i}n},
  {Betancort-Rijo}, {Munari}, {Sartoris}, \& {Barrena}}]{Ferragamo20}
{Ferragamo}, A., {Rubi{\~n}o-Mart{\'i}n}, J.~A., {Betancort-Rijo}, J., {et~al.}
  2020, \aap, 641, A41

\bibitem[{{Hartley} {et~al.}(2008){Hartley}, {Gazzola}, {Pearce}, {Kay}, \&
  {Thomas}}]{hartley08}
{Hartley}, W.~G., {Gazzola}, L., {Pearce}, F.~R., {Kay}, S.~T., \& {Thomas},
  P.~A. 2008, \mnras, 386, 2015

\bibitem[{{Hasselfield} {et~al.}(2013){Hasselfield}, {Hilton}, {Marriage},
  {Addison}, {Barrientos}, {Battaglia}, {Battistelli}, {Bond}, {Crichton},
  {Das}, {Devlin}, {Dicker}, {Dunkley}, {D{\"u}nner}, {Fowler}, {Gralla},
  {Hajian}, {Halpern}, {Hincks}, {Hlozek}, {Hughes}, {Infante}, {Irwin},
  {Kosowsky}, {Marsden}, {Menanteau}, {Moodley}, {Niemack}, {Nolta}, {Page},
  {Partridge}, {Reese}, {Schmitt}, {Sehgal}, {Sherwin}, {Sievers}, {Sif{\'o}n},
  {Spergel}, {Staggs}, {Swetz}, {Switzer}, {Thornton}, {Trac}, \&
  {Wollack}}]{hasselfield13}
{Hasselfield}, M., {Hilton}, M., {Marriage}, T.~A., {et~al.} 2013, Journal of
  Cosmology and Astro-Particle Physics, 2013, 008

\bibitem[{{Heymans} {et~al.}(2012){Heymans}, {Van Waerbeke}, {Miller}, {Erben},
  {Hildebrandt}, {Hoekstra}, {Kitching}, {Mellier}, {Simon}, {Bonnett},
  {Coupon}, {Fu}, {Harnois D{\'e}raps}, {Hudson}, {Kilbinger}, {Kuijken},
  {Rowe}, {Schrabback}, {Semboloni}, {van Uitert}, {Vafaei}, \&
  {Velander}}]{heymans12}
{Heymans}, C., {Van Waerbeke}, L., {Miller}, L., {et~al.} 2012, \mnras, 427,
  146

\bibitem[{{Hildebrandt} {et~al.}(2016){Hildebrandt}, {Choi}, {Heymans},
  {Blake}, {Erben}, {Miller}, {Nakajima}, {van Waerbeke}, {Viola},
  {Buddendiek}, {Harnois-D{\'e}raps}, {Hojjati}, {Joachimi}, {Joudaki},
  {Kitching}, {Wolf}, {Gwyn}, {Johnson}, {Kuijken}, {Sheikhbahaee}, {Tudorica},
  \& {Yee}}]{hildebrandt16}
{Hildebrandt}, H., {Choi}, A., {Heymans}, C., {et~al.} 2016, \mnras, 463, 635

\bibitem[{{Hilton} {et~al.}(2018){Hilton}, {Hasselfield}, {Sif{\'o}n},
  {Battaglia}, {Aiola}, {Bharadwaj}, {Bond}, {Choi}, {Crichton}, {Datta},
  {Devlin}, {Dunkley}, {D{\"u}nner}, {Gallardo}, {Gralla}, {Hincks}, {Ho},
  {Hubmayr}, {Huffenberger}, {Hughes}, {Koopman}, {Kosowsky}, {Louis},
  {Madhavacheril}, {Marriage}, {Maurin}, {McMahon}, {Miyatake}, {Moodley},
  {N{\ae}ss}, {Nati}, {Newburgh}, {Niemack}, {Oguri}, {Page}, {Partridge},
  {Schmitt}, {Sievers}, {Spergel}, {Staggs}, {Trac}, {van Engelen},
  {Vavagiakis}, \& {Wollack}}]{hilton18}
{Hilton}, M., {Hasselfield}, M., {Sif{\'o}n}, C., {et~al.} 2018, The
  Astrophysical Journal Supplement Series, 235, 20

\bibitem[{{Hoekstra} {et~al.}(2015){Hoekstra}, {Herbonnet}, {Muzzin}, {Babul},
  {Mahdavi}, {Viola}, \& {Cacciato}}]{hoekstra15}
{Hoekstra}, H., {Herbonnet}, R., {Muzzin}, A., {et~al.} 2015, \mnras, 449, 685

\bibitem[{{Hook} {et~al.}(2004){Hook}, {J{\o}rgensen}, {Allington-Smith},
  {Davies}, {Metcalfe}, {Murowinski}, \& {Crampton}}]{Hook04}
{Hook}, I.~M., {J{\o}rgensen}, I., {Allington-Smith}, J.~R., {et~al.} 2004,
  \pasp, 116, 425

\bibitem[{{Kay} {et~al.}(2012){Kay}, {Peel}, {Short}, {Thomas}, {Young},
  {Battye}, {Liddle}, \& {Pearce}}]{kay12}
{Kay}, S.~T., {Peel}, M.~W., {Short}, C.~J., {et~al.} 2012, \mnras, 422, 1999

\bibitem[{{Kelly}(2007)}]{kelly07}
{Kelly}, B.~C. 2007, \apj, 665, 1489

\bibitem[{{Komatsu} {et~al.}(2011){Komatsu}, {Smith}, {Dunkley}, {Bennett},
  {Gold}, {Hinshaw}, {Jarosik}, {Larson}, {Nolta}, {Page}, {Spergel},
  {Halpern}, {Hill}, {Kogut}, {Limon}, {Meyer}, {Odegard}, {Tucker}, {Weiland},
  {Wollack}, \& {Wright}}]{Komatsu11_wmap7}
{Komatsu}, E., {Smith}, K.~M., {Dunkley}, J., {et~al.} 2011, \apjs, 192, 18

\bibitem[{{Krause} {et~al.}(2012){Krause}, {Pierpaoli}, {Dolag}, \&
  {Borgani}}]{krause12}
{Krause}, E., {Pierpaoli}, E., {Dolag}, K., \& {Borgani}, S. 2012, \mnras, 419,
  1766

\bibitem[{{{\L}okas} \& {Mamon}(2001)}]{lokas01}
{{\L}okas}, E.~L. \& {Mamon}, G.~A. 2001, \mnras, 321, 155

\bibitem[{{Mamon} {et~al.}(2010){Mamon}, {Biviano}, \& {Murante}}]{Mamon10}
{Mamon}, G.~A., {Biviano}, A., \& {Murante}, G. 2010, \aap, 520, A30

\bibitem[{{Mantz} {et~al.}(2010){Mantz}, {Allen}, {Rapetti}, \&
  {Ebeling}}]{mantz10}
{Mantz}, A., {Allen}, S.~W., {Rapetti}, D., \& {Ebeling}, H. 2010, \mnras, 406,
  1759

\bibitem[{{Medezinski} {et~al.}(2018){Medezinski}, {Battaglia}, {Umetsu},
  {Oguri}, {Miyatake}, {Nishizawa}, {Sif{\'o}n}, {Spergel}, {Chiu}, {Lin},
  {Bahcall}, \& {Komiyama}}]{medezinski18}
{Medezinski}, E., {Battaglia}, N., {Umetsu}, K., {et~al.} 2018, Publications of
  the Astronomical Society of Japan, 70, S28

\bibitem[{{Miyatake} {et~al.}(2019){Miyatake}, {Battaglia}, {Hilton},
  {Medezinski}, {Nishizawa}, {More}, {Aiola}, {Bahcall}, {Bond}, {Calabrese},
  {Choi}, {Devlin}, {Dunkley}, {Dunner}, {Fuzia}, {Gallardo}, {Gralla},
  {Hasselfield}, {Halpern}, {Hikage}, {Hill}, {Hincks}, {Hlo{\v{z}}ek},
  {Huffenberger}, {Hughes}, {Koopman}, {Kosowsky}, {Louis}, {Madhavacheril},
  {McMahon}, {Mandelbaum}, {Marriage}, {Maurin}, {Miyazaki}, {Moodley},
  {Murata}, {Naess}, {Newburgh}, {Niemack}, {Nishimichi}, {Okabe}, {Oguri},
  {Osato}, {Page}, {Partridge}, {Robertson}, {Sehgal}, {Sherwin}, {Shirasaki},
  {Sievers}, {Sif{\'o}n}, {Simon}, {Spergel}, {Staggs}, {Stein}, {Takada},
  {Trac}, {Umetsu}, {van Engelen}, \& {Wollack}}]{miyatake18}
{Miyatake}, H., {Battaglia}, N., {Hilton}, M., {et~al.} 2019, \apj, 875, 63

\bibitem[{{Munari} {et~al.}(2013){Munari}, {Biviano}, {Borgani}, {Murante}, \&
  {Fabjan}}]{munari13}
{Munari}, E., {Biviano}, A., {Borgani}, S., {Murante}, G., \& {Fabjan}, D.
  2013, \mnras, 430, 2638

\bibitem[{{Nagai}(2006)}]{nagai06}
{Nagai}, D. 2006, \apj, 650, 538

\bibitem[{{Navarro} {et~al.}(1997){Navarro}, {Frenk}, \& {White}}]{NFW}
{Navarro}, J.~F., {Frenk}, C.~S., \& {White}, S. D.~M. 1997, \apj, 490, 493

\bibitem[{{Okabe} \& {Smith}(2016)}]{okabe16}
{Okabe}, N. \& {Smith}, G.~P. 2016, \mnras, 461, 3794

\bibitem[{{Penna-Lima} {et~al.}(2017){Penna-Lima}, {Bartlett}, {Rozo}, {Melin},
  {Merten}, {Evrard}, {Postman}, \& {Rykoff}}]{penna17}
{Penna-Lima}, M., {Bartlett}, J.~G., {Rozo}, E., {et~al.} 2017, \aap, 604, A89

\bibitem[{{Planck Collaboration Int.~XXXVI}(2016)}]{nostro16}
{Planck Collaboration Int.~XXXVI}. 2016, \aap, 586, A139

\bibitem[{{Planck Collaboration XX}(2014)}]{planck13_count}
{Planck Collaboration XX}. 2014, \aap, 571, A20

\bibitem[{{Planck Collaboration XXIV}(2016)}]{planck15_count}
{Planck Collaboration XXIV}. 2016, \aap, 594, A24

\bibitem[{{\sorthelp{Planck Collaboration 2014P}}{Planck Collaboration
  XVI}(2014)}]{planck13_par}
{\sorthelp{Planck Collaboration 2014P}}{Planck Collaboration XVI}. 2014, \aap,
  571, A16

\bibitem[{{\sorthelp{Planck Collaboration 2014ZD}}{Planck Collaboration
  XXIX}(2014)}]{planck13_cat}
{\sorthelp{Planck Collaboration 2014ZD}}{Planck Collaboration XXIX}. 2014,
  \aap, 571, A29

\bibitem[{{\sorthelp{Planck Collaboration 2014ZG}}{Planck Collaboration
  XXXII}(2015)}]{planck13_cat_2}
{\sorthelp{Planck Collaboration 2014ZG}}{Planck Collaboration XXXII}. 2015,
  \aap, 581, A14

\bibitem[{{\sorthelp{Planck Collaboration 2015ZB}}{Planck Collaboration
  XXVII}(2016)}]{planck15_cat}
{\sorthelp{Planck Collaboration 2015ZB}}{Planck Collaboration XXVII}. 2016,
  \aap, 594, A27

\bibitem[{{Postman} {et~al.}(2012){Postman}, {Coe}, {Ben{\'\i}tez}, {Bradley},
  {Broadhurst}, {Donahue}, {Ford}, {Graur}, {Graves}, {Jouvel}, {Koekemoer},
  {Lemze}, {Medezinski}, {Molino}, {Moustakas}, {Ogaz}, {Riess}, {Rodney},
  {Rosati}, {Umetsu}, {Zheng}, {Zitrin}, {Bartelmann}, {Bouwens}, {Czakon},
  {Golwala}, {Host}, {Infante}, {Jha}, {Jimenez-Teja}, {Kelson}, {Lahav},
  {Lazkoz}, {Maoz}, {McCully}, {Melchior}, {Meneghetti}, {Merten}, {Moustakas},
  {Nonino}, {Patel}, {Reg{\"o}s}, {Sayers}, {Seitz}, \& {Van der
  Wel}}]{postman12}
{Postman}, M., {Coe}, D., {Ben{\'\i}tez}, N., {et~al.} 2012, The Astrophysical
  Journal Supplement Series, 199, 25

\bibitem[{{Pratt} {et~al.}(2019){Pratt}, {Arnaud}, {Biviano}, {Eckert},
  {Ettori}, {Nagai}, {Okabe}, \& {Reiprich}}]{Pratt19}
{Pratt}, G.~W., {Arnaud}, M., {Biviano}, A., {et~al.} 2019, \ssr, 215, 25

\bibitem[{{Reichardt} {et~al.}(2013){Reichardt}, {Stalder}, {Bleem}, {Montroy},
  {Aird}, {Andersson}, {Armstrong}, {Ashby}, {Bautz}, {Bayliss}, {Bazin},
  {Benson}, {Brodwin}, {Carlstrom}, {Chang}, {Cho}, {Clocchiatti}, {Crawford},
  {Crites}, {de Haan}, {Desai}, {Dobbs}, {Dudley}, {Foley}, {Forman}, {George},
  {Gladders}, {Gonzalez}, {Halverson}, {Harrington}, {High}, {Holder},
  {Holzapfel}, {Hoover}, {Hrubes}, {Jones}, {Joy}, {Keisler}, {Knox}, {Lee},
  {Leitch}, {Liu}, {Lueker}, {Luong-Van}, {Mantz}, {Marrone}, {McDonald},
  {McMahon}, {Mehl}, {Meyer}, {Mocanu}, {Mohr}, {Murray}, {Natoli}, {Padin},
  {Plagge}, {Pryke}, {Rest}, {Ruel}, {Ruhl}, {Saliwanchik}, {Saro}, {Sayre},
  {Schaffer}, {Shaw}, {Shirokoff}, {Song}, {Spieler}, {Staniszewski}, {Stark},
  {Story}, {Stubbs}, {{\v{S}}uhada}, {van Engelen}, {Vanderlinde}, {Vieira},
  {Vikhlinin}, {Williamson}, {Zahn}, \& {Zenteno}}]{Reichardt13}
{Reichardt}, C.~L., {Stalder}, B., {Bleem}, L.~E., {et~al.} 2013, \apj, 763,
  127

\bibitem[{{Ruel} {et~al.}(2014){Ruel}, {Bazin}, {Bayliss}, {Brodwin}, {Foley},
  {Stalder}, {Aird}, {Armstrong}, {Ashby}, {Bautz}, {Benson}, {Bleem},
  {Bocquet}, {Carlstrom}, {Chang}, {Chapman}, {Cho}, {Clocchiatti}, {Crawford},
  {Crites}, {de Haan}, {Desai}, {Dobbs}, {Dudley}, {Forman}, {George},
  {Gladders}, {Gonzalez}, {Halverson}, {Harrington}, {High}, {Holder},
  {Holzapfel}, {Hrubes}, {Jones}, {Joy}, {Keisler}, {Knox}, {Lee}, {Leitch},
  {Liu}, {Lueker}, {Luong-Van}, {Mantz}, {Marrone}, {McDonald}, {McMahon},
  {Mehl}, {Meyer}, {Mocanu}, {Mohr}, {Montroy}, {Murray}, {Natoli},
  {Nurgaliev}, {Padin}, {Plagge}, {Pryke}, {Reichardt}, {Rest}, {Ruhl},
  {Saliwanchik}, {Saro}, {Sayre}, {Schaffer}, {Shaw}, {Shirokoff}, {Song}, {{\v
  S}uhada}, {Spieler}, {Stanford}, {Staniszewski}, {Starsk}, {Story}, {Stubbs},
  {van Engelen}, {Vanderlinde}, {Vieira}, {Vikhlinin}, {Williamson}, {Zahn}, \&
  {Zenteno}}]{ruel14}
{Ruel}, J., {Bazin}, G., {Bayliss}, M., {et~al.} 2014, \apj, 792, 45

\bibitem[{{Saro} {et~al.}(2013){Saro}, {Mohr}, {Bazin}, \& {Dolag}}]{saro13}
{Saro}, A., {Mohr}, J.~J., {Bazin}, G., \& {Dolag}, K. 2013, \apj, 772, 47

\bibitem[{{Sehgal} {et~al.}(2010){Sehgal}, {Bode}, {Das},
  {Hernandez-Monteagudo}, {Huffenberger}, {Lin}, {Ostriker}, \&
  {Trac}}]{sehgal10}
{Sehgal}, N., {Bode}, P., {Das}, S., {et~al.} 2010, \apj, 709, 920

\bibitem[{{Sembolini} {et~al.}(2013){Sembolini}, {Yepes}, {De Petris},
  {Gottl{\"o}ber}, {Lamagna}, \& {Comis}}]{sembolini13}
{Sembolini}, F., {Yepes}, G., {De Petris}, M., {et~al.} 2013, \mnras, 434, 2718

\bibitem[{{Sereno} {et~al.}(2017){Sereno}, {Covone}, {Izzo}, {Ettori},
  {Coupon}, \& {Lieu}}]{sereno17}
{Sereno}, M., {Covone}, G., {Izzo}, L., {et~al.} 2017, \mnras, 472, 1946

\bibitem[{{Sereno} \& {Ettori}(2015)}]{sereno15}
{Sereno}, M. \& {Ettori}, S. 2015, \mnras, 450, 3633

\bibitem[{{Sif{\'o}n} {et~al.}(2016){Sif{\'o}n}, {Battaglia}, {Hasselfield},
  {Menanteau}, {Barrientos}, {Bond}, {Crichton}, {Devlin}, {D{\"u}nner},
  {Hilton}, {Hincks}, {Hlozek}, {Huffenberger}, {Hughes}, {Infante},
  {Kosowsky}, {Marsden}, {Marriage}, {Moodley}, {Niemack}, {Page}, {Spergel},
  {Staggs}, {Trac}, \& {Wollack}}]{sifon16}
{Sif{\'o}n}, C., {Battaglia}, N., {Hasselfield}, M., {et~al.} 2016, \mnras,
  461, 248

\bibitem[{{Smith} {et~al.}(2016){Smith}, {Mazzotta}, {Okabe}, {Ziparo},
  {Mulroy}, {Babul}, {Finoguenov}, {McCarthy}, {Lieu}, {Bah{\'e}}, {Bourdin},
  {Evrard}, {Futamase}, {Haines}, {Jauzac}, {Marrone}, {Martino}, {May},
  {Taylor}, \& {Umetsu}}]{smith16}
{Smith}, G.~P., {Mazzotta}, P., {Okabe}, N., {et~al.} 2016, \mnras, 456, L74

\bibitem[{{Springel}(2005)}]{springel05}
{Springel}, V. 2005, \mnras, 364, 1105

\bibitem[{{Stanek} {et~al.}(2010){Stanek}, {Rasia}, {Evrard}, {Pearce}, \&
  {Gazzola}}]{stanek10}
{Stanek}, R., {Rasia}, E., {Evrard}, A.~E., {Pearce}, F., \& {Gazzola}, L.
  2010, \apj, 715, 1508

\bibitem[{{Sunyaev} \& {Zeldovich}(1972)}]{SZ72}
{Sunyaev}, R.~A. \& {Zeldovich}, Y.~B. 1972, Comments on Astrophysics and Space
  Physics, 4, 173

\bibitem[{{Tinker} {et~al.}(2008){Tinker}, {Kravtsov}, {Klypin}, {Abazajian},
  {Warren}, {Yepes}, {Gottl{\"o}ber}, \& {Holz}}]{tinker08}
{Tinker}, J., {Kravtsov}, A.~V., {Klypin}, A., {et~al.} 2008, \apj, 688, 709

\bibitem[{{Tremaine} {et~al.}(2002){Tremaine}, {Gebhardt}, {Bender}, {Bower},
  {Dressler}, {Faber}, {Filippenko}, {Green}, {Grillmair}, {Ho}, {Kormendy},
  {Lauer}, {Magorrian}, {Pinkney}, \& {Richstone}}]{Tremaine02}
{Tremaine}, S., {Gebhardt}, K., {Bender}, R., {et~al.} 2002, \apj, 574, 740

\bibitem[{{van der Burg} {et~al.}(2016){van der Burg}, {Aussel}, {Pratt},
  {Arnaud}, {Melin}, {Aghanim}, {Barrena}, {Dahle}, {Douspis}, {Ferragamo},
  {Fromenteau}, {Herbonnet}, {Hurier}, {Pointecouteau},
  {Rubi{\~n}o-Mart{\'{\i}}n}, \& {Streblyanska}}]{remco16}
{van der Burg}, R.~F.~J., {Aussel}, H., {Pratt}, G.~W., {et~al.} 2016, \aap,
  587, A23

\bibitem[{{Vanderlinde} {et~al.}(2010){Vanderlinde}, {Crawford}, {de Haan},
  {Dudley}, {Shaw}, {Ade}, {Aird}, {Benson}, {Bleem}, {Brodwin}, {Carlstrom},
  {Chang}, {Crites}, {Desai}, {Dobbs}, {Foley}, {George}, {Gladders}, {Hall},
  {Halverson}, {High}, {Holder}, {Holzapfel}, {Hrubes}, {Joy}, {Keisler},
  {Knox}, {Lee}, {Leitch}, {Loehr}, {Lueker}, {Marrone}, {McMahon}, {Mehl},
  {Meyer}, {Mohr}, {Montroy}, {Ngeow}, {Padin}, {Plagge}, {Pryke}, {Reichardt},
  {Rest}, {Ruel}, {Ruhl}, {Schaffer}, {Shirokoff}, {Song}, {Spieler},
  {Stalder}, {Staniszewski}, {Stark}, {Stubbs}, {van Engelen}, {Vieira},
  {Williamson}, {Yang}, {Zahn}, \& {Zenteno}}]{Vanderlinde10}
{Vanderlinde}, K., {Crawford}, T.~M., {de Haan}, T., {et~al.} 2010, \apj, 722,
  1180

\bibitem[{{Vikhlinin} {et~al.}(2009){Vikhlinin}, {Burenin}, {Ebeling},
  {Forman}, {Hornstrup}, {Jones}, {Kravtsov}, {Murray}, {Nagai}, {Quintana}, \&
  {Voevodkin}}]{vikhlinin09}
{Vikhlinin}, A., {Burenin}, R.~A., {Ebeling}, H., {et~al.} 2009, \apj, 692,
  1033

\bibitem[{{von der Linden} {et~al.}(2014){von der Linden}, {Mantz}, {Allen},
  {Applegate}, {Kelly}, {Morris}, {Wright}, {Allen}, {Burchat}, {Burke},
  {Donovan}, \& {Ebeling}}]{von_der_Linden14}
{von der Linden}, A., {Mantz}, A., {Allen}, S.~W., {et~al.} 2014, \mnras, 443,
  1973

\bibitem[{{Wainer} \& {Thissen}(1976)}]{gapper}
{Wainer}, H. \& {Thissen}, D. 1976, Psychometrika, 41, 9

\bibitem[{{Williamson} {et~al.}(2011){Williamson}, {Benson}, {High}, {Vand
  erlinde}, {Ade}, {Aird}, {Andersson}, {Armstrong}, {Ashby}, {Bautz}, {Bazin},
  {Bertin}, {Bleem}, {Bonamente}, {Brodwin}, {Carlstrom}, {Chang}, {Chapman},
  {Clocchiatti}, {Crawford}, {Crites}, {de Haan}, {Desai}, {Dobbs}, {Dudley},
  {Fazio}, {Foley}, {Forman}, {Garmire}, {George}, {Gladders}, {Gonzalez},
  {Halverson}, {Holder}, {Holzapfel}, {Hoover}, {Hrubes}, {Jones}, {Joy},
  {Keisler}, {Knox}, {Lee}, {Leitch}, {Lueker}, {Luong-Van}, {Marrone},
  {McMahon}, {Mehl}, {Meyer}, {Mohr}, {Montroy}, {Murray}, {Padin}, {Plagge},
  {Pryke}, {Reichardt}, {Rest}, {Ruel}, {Ruhl}, {Saliwanchik}, {Saro},
  {Schaffer}, {Shaw}, {Shirokoff}, {Song}, {Spieler}, {Stalder}, {Stanford},
  {Staniszewski}, {Stark}, {Story}, {Stubbs}, {Vieira}, {Vikhlinin}, \&
  {Zenteno}}]{Williamson11}
{Williamson}, R., {Benson}, B.~A., {High}, F.~W., {et~al.} 2011, \apj, 738, 139

\bibitem[{{Yang} {et~al.}(2010){Yang}, {Bhattacharya}, \& {Ricker}}]{yang10}
{Yang}, H. Y.~K., {Bhattacharya}, S., \& {Ricker}, P.~M. 2010, \apj, 725, 1124

\end{thebibliography}

\appendix
\section{Tables}
\label{app:A}
In this appendix we present all the tables described in Sec.~\ref{sec:ITP_SDSS_sample}. The optical coordinates, the distance from the \planck{} ponting, the redshift, and the number of galaxy members (col. 4-8) were previously given in  \cite{nostro16} and \cite{rafa18, rafa20}. We report the velocity dispersion and the dynamical mass of these GCs (col.9-10) for the first time. All these tables are also available in electronic format.
\onecolumn

\begin{landscape}
\label{tab:vd_others_flag1}
\setlength\LTleft{0pt}
\setlength\LTright{0pt}
\tiny

\end{longtabu}
\end{landscape}

\twocolumn

\end{document}